\documentclass{ws-ijmpd}
\usepackage[super,compress]{cite}
\usepackage{longtable}
\begin{document}

\markboth{Scarpa et al.}
{Testing Newtonian dynamics with wide binary stars}

%
\catchline{}{}{}{}{}
%

\title{Dynamics of wide binary stars: A case study for testing 
Newtonian dynamics in the low acceleration regime}

\author{Riccardo Scarpa}
\address{Instituto de Astrof\'isica de Canarias, c/via Lactea s/n\\
San Cristobal de la Laguna, 38205, Spain\\
Departamento de Astrofísica, Universidad de La Laguna (ULL), 38206 
La Laguna, Tenerife, Spain\\
riccardo.scarpa@gtc.iac.es}

\author{Riccardo Ottolina}
\address{Dipartimento di Scienza e Alta Tecnologia, Universit\`a degli 
Studi dell'Insubria, via Valleggio 11\\
 Como, 22100, Italy \\
}

\author{Renato Falomo}
\address{Osservatorio Astronomico di Padova, INAF, vicolo 
dell' Osservatorio 5\\ Padova, 35122, Italy\\
renato.falomo@oapd.inaf.it}

\author{Aldo Treves}
\address{Dipartimento di Scienza e Alta Tecnologia, Universit\`a degli 
Studi dell'Insubria, via Valleggio 11\\
 Como, 22100, Italy \\
aldo.treves@uninsubria.it}
\maketitle

\begin{history}
\received{Day Month Year}
\revised{Day Month Year}
\end{history}

\abstract Extremely wide binary stars represent ideal systems to probe
Newtonian dynamics in the low acceleration regimes ($<$ 10$^{-10}$ m
$s^{-2}$) typical of the external regions of galaxies.  Here we
present a study of 60 alleged wide binary stars with projected
separation ranging from 0.004 to 1 pc, probing gravitational
accelerations well below the limit were dark matter or modified
dynamics theories set in.  Radial velocities with accuracy $\sim 100$
m/s were obtained for each star, in order to constrain their orbital
velocity, that, together with proper motion data, can distinguish
bound from unbound systems.  It was found that about half of the
observed pairs do have velocity in the expected range for bound
systems, out to the largest separations probed here. In particular, we
identified five pairs with projected separation $>0.15$ pc that are
useful for the proposed test.  While it would be premature to draw any
conclusion about the validity of Newtonian dynamics at these low
accelerations, our main result is that very wide binary stars seem to
exist in the harsh environment of the solar neighborhood. This could
provide a tool to test Newtonian dynamics versus modified dynamics
theories in the low acceleration conditions typical of galaxies. In
the near future the GAIA satellite will provide data to increase
significantly the number of wide pairs that, with the appropriate
follow up spectroscopic observations, will allow the implementation of
this experiment with unprecedented accuracy.


\ccode{PACS numbers:  04.50.Kd, 95.35.+d, 97.10.Vm,  97.80.-d, }


\section{Introduction}

A wealth of astronomical data indicate the dynamics in a variety of
classes of cosmic structures deviate from the expectation of Newtonian
dynamics applied to the mass visible in stars and gas. These
observations are usually explained invoking the existence of vast
amounts of unseen mass in some novel form, dark matter
(e.g. Refs.~\refcite{sanders10}, \refcite{bertone16}) .  Alternatively
the data could be interpreted as a breakdown of Newtonian dynamics 
on the relevant regime.  While most investigators would agree
that the former is the correct explanation, an increasing number of
empirical results support the latter (e.g. Ref.~\refcite{famaey11} for
a recent review).  In particular, the proliferation of alternative
theories of dynamics found nowadays in the literature points to the
presence of an acceleration scale --- $a_0 \sim 10^{-10}$ m s$^{-2}$
--- above which classical dynamics is recovered and below which the
dark matter mimicking regime appears.

The vast majority of the work on alternative theories of gravity was
focused on explaining the dynamical properties of galaxies, most
notably the detailed shape their flat 
rotation curve, {e.g. Ref.~\refcite{sanders02}}.

When modifying gravity in the low-acceleration regime, the strong
equivalence principle is in general broken, and systems embedded in a
stronger field than their internal gravitational field might not
display a modified gravity behaviour. However, this is not the case
for any possible theory of modified dynamics, as shown by, e.g.,
\refcite{milgrom11}): it is thus conceivable that the modified
dynamics regime kicks in below a given acceleration scale
independently of the external field. According to this, in the past
years we tested Newtonian dynamics using stars
in globular clusters (See
Refs.~\refcite{scarpa11}-\refcite{scarpa03}), which due to their small
size are believed to contain negligible amount of dark matter, if any.
It was found that the orbital velocities of stars in their outskirt,
where $a<a_0$, are too large to be consistent with Newtonian dynamics.
A result that has been corroborated for a growing number of globular
clusters by various independent groups
(e.g. Refs.~\refcite{lane09},\refcite{lane11}).  The interpretation of
this result is, however, complicated by a number of effects that could
mimic the modified dynamics behavior --- cluster evaporation, tidal heating,
peculiar orbital motion, uncertain mass-to-light ratio, dark remmants
--- thus preventing to draw a clear cut conclusion.

In order to carry out a cleaner test, free from contaminating external
effects, one might look at the simplest stellar system: wide binary
stars. Indeed, sufficiently wide binaries could probe the acceleration
regime typical of galaxies, thus permitting a direct test of Newtonian
dynamics below $a_0$.  

For a test particle orbiting around a 1M$_\odot$ star, modified
gravity supposedly sets in at separation of $r_0\sim$7000 AU (0.03
pc), and the expected orbital velocity is $\sim$250 m/s.  This
separation is so large and the orbital velocity so low that at the
time of writing, there is no hope to directly trace the orbital
motion.  Indeed, it is not even clear whether such wide binaries can
exist at all in the harsh environment nearby the Sun, where tidal
effects and close encounter with other stars could destroy them (see
e.g. Ref.~\refcite{jiang10}).

Assuming that such wide binaries exist and have a stable orbit --- a
hypothesis corroborated by detailed numerical simulations showing that
the effects on wide binaries of the Milky Way external field do not
alter the Keplerian fall-off at least up to separations of the Jacobi
radius, which in the solar vicinity is $\sim$ 1.7 pc (Ref.~\refcite{jiang10})
--- the gravitational acceleration can be constrained measuring the
instantaneous orbital velocity of the two components.  Considering a
large sample of binaries, uniformly covering a wide range of
separations, it is then possible to trace the run of velocity with
distance, in this way building the equivalent of a galaxy rotation
curve.  Compared to other experiments, the use of binary stars has the
major advantage that the masses involved are known from their spectral
type, thus removing one important source of uncertainty (the
mass-to-light ratio), making this test one of the cleanest possible.

Therefore since double stars are not surrounded by an halo of dark
matter, Newtonian dynamics predicts this ``rotation curve'' must fall
off as $v\propto r^{-1/2}$, essentially following Kepler's third law,
provided the range of masses is small. On the contrary, if the
velocity converges toward a relatively large, constant value as seen in
galaxies, then we will be forced to seriously question our
understanding of Newton's law of gravity in low acceleration regimes.

In the case only one or two components of the velocity vector is
known, projection effects play an important role and the Newtonian
predictions become an upper limit to the observed velocity.  A
preliminary attempt to carry out a test of this kind was performed by
Hernandez et al. (Ref.~\refcite{Hernandez12}).  They discuss the
difference of orbital velocity derived from Hipparcos proper motion
for a large sample of binaries, suggesting that the orbital velocity
might $not$ decrease with the star separation. Unfortunately, the data
discussed by Hernandez et al. (Ref.~\refcite{Hernandez12}) have large
uncertainties (average error on velocity 0.8 km/s) making their result
rather weak and inconclusive.  Moreover, not having information about
the radial velocity of these stars, the proper motion data alone leave
open the possibility that a significant fraction of the stars under
consideration are not gravitationally bound even though the proper
motion data suggest so.

Here we present results of a first attempt to cope with this last
problem.  A number of alleged wide double stars --- selected mostly
according to Hipparcos parallaxes and proper motion data --- were
observed spectroscopically to derive their radial velocity
with accuracy of $\sim 100$ m s$^{-1}$ so to build their 3d velocity
vector and to confirm the velocity difference of the two components is
small, as it should be for bound systems. As we will see, while most
alleged doubles are not confirmed as such by the new data, a number of
them are consistent with the hypothesis of being bound systems, making
this test feasible.

\section{What are the expectations?}

\subsection{The case of standard Newtonian dynamics}

Let assume a double star with components of mass $m_1$ and $m_2$, separated
by $S = r_1 + r_2$, where $r_1$ and $r_2$ are the distances of each
star from the center of mass of the system. Equating the momentum $m_1 r_1 = m_2 r_2$
of the two masses with respect to the center of mass
we obtain:
\begin{equation}
r_1 = \frac{S m_2}{m_1+m_2}
\end{equation}
and a similar expression for $r_2$.
Assuming for simplicity circular orbit, equating centripetal acceleration
to gravitational acceleration on $m_1$: 
\begin{equation}
\frac{v_1^2}{r_1} = \frac{Gm_2}{S^2}
\end{equation}
where $v_1$ is the velocity of component 1, which becomes:
\begin{equation}
v_1^2 = \frac{Gm_2^2}{S(m_1+m_2)}.
\end{equation}
and a similar expression for $v_2$. In the simplest case 
where $m_1=m_2=m$ and consequently $v_1=v_2=v$ we further get
\begin{equation}
v^2 = \frac{Gm}{2S}.
\end{equation}
This last expression gives the expected orbital velocity for each star. 
However, because the two components of a pair are moving in opposite directions, the 
$observed$ velocity difference $\Delta V$ is twice as large. 
Thus, calling $m_{tot} = 2m$
the total mass of the system, we finally have:
\begin{equation}
\Delta V_{observed} = \sqrt{\frac{Gm_{tot}}{S}}
\end{equation}
The resulting orbital velocity for two stars of 1M$_\odot$ each and
separation $S=7000$ AU  is 250 m/s, with a corresponding $observed$ $\Delta
V \sim 500$ m/s.
This value corresponds to the 3d velocity vector. When considering
only the radial velocity, projection effects and orbital ellipticity
will play a role, making this an upper limit to the radial velocity
difference we should expect to observe. Whatever the case, the orbital velocity
should decrease with separation.
 
\subsection{Expectations for modified dynamics }

An alternative approach to the missing mass problem is to replace dark
matter by a modified dynamics theory.  In the literature various
approaches have been proposed.  The best known of them being the
MOdified Newtonian Dynamics (MOND) or its relativistic version called
Tensor–Vector–Scalar (TeVeS) theory (Refs.~\refcite{milgrom83a} -
\refcite{Bekenstein04}). Another example is the generally covariant
MOdified Gravity (MOG) theory (Ref.~\refcite{moffat06}).  Some
successfull attempt to apply quantum gravity to galaxies were also made
(Ref.~\refcite{Rodrigues10}), or by introducing non-local gravity
(Ref.~\refcite{Hehl09}).

Here, to quantify the effects of modified dynamics we adopt, for its
semplicity, the MOND working formula with the standard interpolation
function (Refs.~\refcite{milgrom83a}-\refcite{milgrom83c}). In doing
so, we assign no special value to MOND compared to the others
theories, in particular because due to the external field effect, MOND
as a modified gravity formula should not be directly applicable to
systems surrounded by a strong external field, as is the case of stars
inside the Milky Way (Ref.~\refcite{milgrom83a}), though the effects
we are looking for in this work could actually be present in some
version of MOND (\refcite{milgrom11}).

According to MOND formula for acceleration of gravity well below 
$a_0 = 1.2 \times 10^{-10}$ m s$^{-2}$, the acceleration should become
\begin{equation}
a = \sqrt{a_Na_0}
\end{equation}
\noindent where $a_N$ is the usual Newtonian acceleration. Therefore 
for the simplest case of a double star with equal mass components and circular orbits 
outlined above we can write:
\begin{equation}
\frac{v}{S} = \sqrt{\frac{Gm_{tot}a_0}{4S^2}}
\end{equation}
\noindent from which one immediately gets
\begin{equation}
v = \left( \frac{Gm_{tot}a_0}{4} \right) ^{1/4}.
\end{equation}
\noindent 

The most important aspect of this relation is that the separation of
the two stars disappeared, so that the orbital velocity becomes
constant. This behavior is similar to what is observed in rotation
curves of galaxies. That is, MOND-like formulae suggest that for
separations larger than $\sim 7000$ AU the orbital velocity should be
constant with asymptotic value of $\sim 300$ m/s in the case of
1M$_\odot$ stars, and a corresponding $observed$ $\Delta V$ twice as
big.  The dependence on the mass is very weak, thus this limit applies
to all double stars considered in this study, which have masses in the
range $0.4<m<1.5$ solar masses. Since in Newtonian dynamics the orbital
velocity keeps decreasing, the difference between the two scenarios
becomes significant at larger separations (lower acceleration).  For
instance, at 0.1 pc separation, the predicted orbital velocity is twice 
the Newtonian value.

\section{ The sample of wide binary stars candidates.}

Wide binary star candidates were selected from the Shaya \& Olling
catalogue (Ref.~\refcite{shaya11}). In this catalog wide binaries are
identified by assigning a probability above chance alignment for each
system, with probability obtained using a sophisticated Bayesian
statistical analysis in a multi-dimensional parameter space of proper
motions and spatial positions of the Hipparcos catalog
(Ref.~\refcite{perryman97}).

From the Shaya \& Olling catalogue we selected isolated binaries with
a probability of non-chance alignment greater than 99\%. The binary
search criteria used by the authors require that the proposed binary
should have no near neighbors; the projected separation between the
two components is thus always many times smaller than the typical
interstellar separation, which is very important for our test.  It is
also important to note the catalogue includes both ``presently bound''
and ``previously bound'' pairs. Previously bound pairs are those which
according to the authors have been destroyed by the Galactic tidal
field and perturbation from nearby stars
(Ref.~\refcite{jiang10}). Considering we are testing a scenario of
modified dynamics, having also these ``previously bound'' stars ensure
we are not missing precisely those systems moving too fast for Newton
but not for modified dynamics.

In summary we select pairs from the catalogue according to the following criteria:
\begin{itemlist}
\item stars are classified as double, not multiple;
\item probability of the pair to be physically bound (or previously bound) greater than  99\%;
\item mass between 0.4 and 1.5 solar masses;
\item pair projected separation smaller than 1.5 pc;
\item declination $> -40^{\circ}$ for good visibility from  Roque de los Muchachos observatory (La Palma);
\item both stars in the pair are not known to be spectroscopic binary;
\item relative error on the distance $<$ 15\%.
\end{itemlist}
\indent This selection yielded 60 pairs of stars (see Table \ref{Tab:1}) in
the magnitude range $4<V<12$, and distances going from 14 pc to 100
pc.  Note that when referring to the distance of a pair we mean the
average distance of the two components.  The 55\% of the sample stars are
within 50 pc with associated errors on distance of few parsecs.
Apparent separations of selected stars range from 30 to 3650 arcsec,
which according to distance correspond to projected physical
separation in the 0.004 to 1.3 pc range, most of the pairs being
within 0.1 pc separation.  As mentioned before we are interested in
the behavior of binaries with separation above $r_0=7000$ AU, that is 0.034
pc, while smaller separations will be used to check whether the
Newtonian dynamics is recovered. The sample includes 29 and 31 pairs
with separation smaller and larger than this value, respectively.
\begin{longtable}{cccccccc}
\caption{The selected sample (horizontal lines separate different pairs).\\
Legend:\\
\textbf{NH} = Hipparcos catalogue star number;\\
\textbf{RA} = Right ascension;\\
\textbf{DEC} = Declination;\\
\textbf{V} = Apparent V band magnitude;\\
\textbf{ST} = Spectral type;\\
\textbf{d} = Distance from parallax;\\
\textbf{pmRA}= Proper motion along RA in milliarcsec/year;\\
\textbf{pmDEC}= Proper motion along DEC in milliarcsec/year.\\
} \label{Tab:1}\\

\hline
NH & RA & DEC & V & ST & d & pmRA & pmDEC \\
 & [hh:mm:ss]  & [dd:mm:ss]  &  & & [pc]   &  [msec/yr] & [msec/yr]\\                     
\hline
\endfirsthead

\multicolumn{8}{l}{{\bfseries \tablename\ \thetable{} -- continued from previous page}} \\
\hline
NH & RA & DEC & V & ST & d & pmRA & pmDEC \\
 & [hh:mm:ss]  & [dd:mm:ss]  &  & & [pc]   &  [msec/yr] & [msec/yr]\\                   
\hline
\endhead

\hline
\multicolumn{8}{r}{ \bfseries Continued on next page}
\endfoot

\hline
\endlastfoot

 185 & 00:02:21.64 & 10:47:08.32 & 8.5 & F8 & 77.4 $ \pm $ 5.5 & -45.33 $ \pm $ 1.61 & -115.73 $ \pm $ 1.09 \\ 
 190 & 00:02:25.33 & 10:46:35.95 & 8.7 & G0 & 87.5 $ \pm $ 7.1 & -49.43 $ \pm $ 1.06 & -118.65 $ \pm $ 0.79 \\ 
\hline                                    
 201 & 00:02:33.44 & 18:41:00.11 & 8.2 & F5 & 92.2 $ \pm $ 8.9 & -13.19 $ \pm $ 0.70 & 17.41 $ \pm $ 0.56 \\ 
 206 & 00:02:35.24 & 18:50:09.58 & 8.6 & G0 & 99.9 $ \pm $ 9.6 & -18.50 $ \pm $ 0.78 & 19.70 $ \pm $ 0.59 \\ 
\hline                                    
 1891 & 00:23:53.20 & 29:30:09.10 & 8.2 & G & 84.6 $ \pm $ 7.5 & 41.80 $ \pm $ 1.18 & -3.47 $ \pm $ 0.82 \\ 
 1887 & 00:23:51.46 & 29:29:45.29 & 8.5 & G & 71.9 $ \pm $ 4.9 & 39.41 $ \pm $ 1.30 & 0.11 $ \pm $ 0.90 \\ 
\hline                                    
 2292 & 00:29:16.18 & -05:54:35.82 & 7.8 & G0 & 56.7 $ \pm $ 2.5 & -113.02 $ \pm $ 1.07 & -220.86 $ \pm $ 0.68 \\ 
 2350 & 00:29:59.93 & -05:45:48.45 & 9.4 & G5 & 49.9 $ \pm $ 3.7 & -107.08 $ \pm $ 1.51 & -224.06 $ \pm $ 0.91 \\ 
\hline                                    
 4702 & 01:00:27.93 & -19:23:21.33 & 8.0 & G3V & 74.7 $ \pm $ 5.1 & 127.66 $ \pm $ 1.02 & -54.14 $ \pm $ 0.67 \\ 
 4833 & 01:02:06.82 & -19:40:10.35 & 8.3 & G3V & 86.9 $ \pm $ 6.5 & 129.62 $ \pm $ 1.04 & -54.00 $ \pm $ 0.69 \\ 
\hline                                    
 8497 & 01:49:35.19 & -10:41:10.25 & 4.7 & F3III & 23.2 $ \pm $ 0.1 & -148.90 $ \pm $ 0.87 & -94.47 $ \pm $ 0.86 \\ 
 8486 & 01:49:23.43 & -10:42:11.93 & 6.7 & G0 & 22.6 $ \pm $ 1.5 & -122.64 $ \pm $ 4.36 & -100.38 $ \pm $ 4.09 \\ 
\hline                                    
 10321 & 02:12:54.96 & 40:40:06.98 & 7.2 & G0V & 26.8 $ \pm $ 0.4 & 59.29 $ \pm $ 0.71 & -108.78 $ \pm $ 0.67 \\ 
 10339 & 02:13:13.29 & 40:30:28.16 & 7.3 & G0V & 26.2 $ \pm $ 0.4 & 58.11 $ \pm $ 0.65 & -95.90 $ \pm $ 0.76 \\ 
\hline                                    
 11137 & 02:23:19.40 & 15:25:05.21 & 8.9 & G5 & 58.8 $ \pm $ 4.6 & 112.85 $ \pm $ 1.68 & 185.66 $ \pm $ 1.11 \\ 
 11134 & 02:23:17.03 & 15:24:59.62 & 9.4 & G5 & 57.1 $ \pm $ 5.4 & 111.79 $ \pm $ 2.09 & 185.42 $ \pm $ 1.33 \\ 
\hline                                    
 11783 & 02:32:05.28 & -15:14:39.55 & 4.7 & F5V & 26.7 $ \pm $ 0.2 & -80.92 $ \pm $ 0.69 & -146.84 $ \pm $ 0.64 \\ 
 11759 & 02:31:42.52 & -15:16:23.39 & 8.7 & K2.5Vk & 27.9 $ \pm $ 0.7 & -75.71 $ \pm $ 1.29 & -120.03 $ \pm $ 1.36 \\ 
\hline                                    
 15304 & 03:17:26.29 & 07:39:20.97 & 7.4 & F8V & 46.1 $ \pm $ 2.0 & 169.30 $ \pm $ 1.66 & -7.64 $ \pm $ 0.96 \\ 
 15310 & 03:17:32.68 & 07:41:24.61 & 7.8 & G0 & 41.8 $ \pm $ 2.1 & 170.41 $ \pm $ 1.93 & -7.48 $ \pm $ 1.10 \\ 
\hline                                    
 15527 & 03:20:03.35 & -28:51:14.09 & 7.4 & G1.5V & 35.5 $ \pm $ 0.8 & 348.88 $ \pm $ 0.50 & -64.82 $ \pm $ 0.73 \\ 
 15526 & 03:20:02.71 & -28:47:01.21 & 8.5 & G9.5V & 35.4 $ \pm $ 1.4 & 349.07 $ \pm $ 0.78 & -67.80 $ \pm $ 1.00 \\ 
\hline                                    
 17118 & 03:39:58.64 & 63:52:14.74 & 6.8 & F5 & 42.5 $ \pm $ 0.9 & 129.42 $ \pm $ 0.58 & -143.11 $ \pm $ 0.66 \\ 
 17126 & 03:40:05.21 & 63:52:29.81 & 8.2 & G5 & 40.7 $ \pm $ 1.7 & 137.58 $ \pm $ 0.94 & -131.99 $ \pm $ 1.04 \\ 
\hline                                    
 19335 & 04:08:36.50 & 38:02:24.82 & 5.5 & F7V & 21.0 $ \pm $ 0.1 & 163.93 $ \pm $ 0.65 & -203.52 $ \pm $ 0.55 \\ 
 19255 & 04:07:34.22 & 38:04:30.31 & 7.1 & G5 & 20.4 $ \pm $ 0.3 & 172.94 $ \pm $ 1.01 & -226.60 $ \pm $ 0.89 \\ 
\hline                                    
 19859 & 04:15:28.86 & 06:11:13.64 & 6.3 & G0IV & 21.3 $ \pm $ 0.2 & -109.37 $ \pm $ 1.12 & -108.35 $ \pm $ 1.12 \\ 
 19855 & 04:15:25.85 & 06:11:59.73 & 6.9 & G5IV & 21.1 $ \pm $ 0.3 & -101.62 $ \pm $ 1.16 & -112.85 $ \pm $ 1.15 \\ 
\hline                                    
 20342 & 04:21:28.68 & -20:54:55.48 & 8.8 & K2.5V & 38.6 $ \pm $ 2.2 & 190.51 $ \pm $ 1.33 & 130.36 $ \pm $ 1.36 \\ 
 20338 & 04:21:26.29 & -20:54:06.56 & 11.5 & K9Vk: & 30.8 $ \pm $ 3.8 & 197.15 $ \pm $ 2.52 & 141.96 $ \pm $ 3.16 \\ 
\hline                                    
 20669 & 04:25:40.26 & 63:40:29.47 & 8.3 & G0 & 62.4 $ \pm $ 2.9 & -128.25 $ \pm $ 0.90 & -55.73 $ \pm $ 0.98 \\ 
 20637 & 04:25:22.84 & 63:37:34.70 & 10.0 & K0 & 64.3 $ \pm $ 6.3 & -128.46 $ \pm $ 1.62 & -51.39 $ \pm $ 1.96 \\ 
\hline                                    
 21537 & 04:37:26.71 & 00:33:11.13 & 7.5 & G5 & 66.9 $ \pm $ 3.2 & 17.23 $ \pm $ 1.20 & 11.42 $ \pm $ 0.95 \\ 
 21534 & 04:37:26.08 & 00:34:28.57 & 7.5 & G5 & 66.1 $ \pm $ 3.4 & 15.21 $ \pm $ 1.17 & 12.90 $ \pm $ 0.95 \\ 
\hline                                    
 21704 & 04:39:37.02 & -21:14:51.36 & 7.2 & K0/K1I & 87.1 $ \pm $ 5.9 & 12.96 $ \pm $ 0.95 & 21.67 $ \pm $ 0.95 \\ 
 21702 & 04:39:34.95 & -21:14:23.14 & 9.0 & G0 & 90.9 $ \pm $ 13.1 & 13.01 $ \pm $ 1.26 & 22.20 $ \pm $ 1.30 \\ 
\hline                                    
 22611 & 04:51:54.19 & -34:14:19.25 & 6.7 & F6IV/V & 59.3 $ \pm $ 1.4 & 75.89 $ \pm $ 0.45 & 9.10 $ \pm $ 0.54 \\ 
 22604 & 04:51:47.90 & -34:13:17.27 & 8.8 & G5V & 57.6 $ \pm $ 2.7 & 73.75 $ \pm $ 0.70 & 7.86 $ \pm $ 0.86 \\ 
\hline                                    
 24046 & 05:10:03.74 & 27:33:24.20 & 7.0 & F8V & 40.0 $ \pm $ 1.3 & 197.69 $ \pm $ 1.11 & -88.54 $ \pm $ 0.59 \\ 
 24035 & 05:10:00.74 & 27:38:36.24 & 9.2 & G5 & 39.7 $ \pm $ 1.9 & 207.69 $ \pm $ 1.54 & -84.28 $ \pm $ 0.80 \\ 
\hline                                    
 25278 & 05:24:25.31 & 17:23:00.79 & 5.0 & F8V & 14.4 $ \pm $ 0.1 & 250.40 $ \pm $ 0.88 & -7.42 $ \pm $ 0.61 \\ 
 25220 & 05:23:38.23 & 17:19:26.87 & 7.9 & K2 & 14.1 $ \pm $ 0.3 & 253.35 $ \pm $ 1.21 & -4.66 $ \pm $ 0.76 \\ 
\hline                                    
 33705 & 07:00:09.76 & -31:08:30.38 & 6.6 & F5.5VF & 37.4 $ \pm $ 0.5 & 19.70 $ \pm $ 0.58 & 34.37 $ \pm $ 0.64 \\ 
 33691 & 07:00:02.73 & -31:13:41.42 & 8.4 & G9V & 37.1 $ \pm $ 1.2 & 20.44 $ \pm $ 0.82 & 35.10 $ \pm $ 0.93 \\ 
\hline                                    
 34426 & 07:08:12.02 & 15:31:16.91 & 7.7 & F8 & 45.9 $ \pm $ 1.6 & -56.97 $ \pm $ 1.04 & -212.48 $ \pm $ 0.81 \\ 
 34407 & 07:08:00.27 & 15:31:44.75 & 7.8 & G0V & 46.8 $ \pm $ 1.9 & -51.42 $ \pm $ 1.07 & -206.54 $ \pm $ 0.84 \\ 
\hline                                    
 34714 & 07:11:19.56 & 33:06:42.78 & 7.2 & F5 & 48.2 $ \pm $ 1.4 & -111.55 $ \pm $ 1.62 & -8.00 $ \pm $ 0.85 \\ 
 34700 & 07:11:14.80 & 32:36:54.12 & 8.0 & G0 & 45.2 $ \pm $ 1.6 & -110.82 $ \pm $ 1.32 & -5.71 $ \pm $ 0.79 \\ 
\hline                                    
 39457 & 08:03:53.84 & -31:32:40.62 & 8.7 & G3V & 73.3 $ \pm $ 6.2 & -74.05 $ \pm $ 0.95 & 148.61 $ \pm $ 1.08 \\ 
 39452 & 08:03:50.71 & -31:33:06.99 & 9.7 & - & 71.3 $ \pm $ 10.0 & -70.51 $ \pm $ 6.29 & 143.46 $ \pm $ 7.40 \\ 
\hline                                    
 40918 & 08:21:03.76 & 65:26:33.58 & 8.0 & G0 & 36.5 $ \pm $ 1.2 & 24.63 $ \pm $ 0.74 & 21.56 $ \pm $ 0.72 \\ 
 40882 & 08:20:33.34 & 65:23:38.11 & 8.4 & G5 & 37.3 $ \pm $ 1.6 & 11.65 $ \pm $ 0.85 & 21.55 $ \pm $ 0.84 \\ 
\hline                                    
 44520 & 09:04:15.04 & 03:01:35.71 & 8.9 & G5 & 56.2 $ \pm $ 4.9 & 58.67 $ \pm $ 1.56 & -92.23 $ \pm $ 0.77 \\ 
 44518 & 09:04:13.87 & 03:02:02.72 & 9.3 & G5 & 47.8 $ \pm $ 4.0 & 48.12 $ \pm $ 1.87 & -87.70 $ \pm $ 0.92 \\ 
\hline                                    
 44858 & 09:08:23.90 & 27:32:06.96 & 8.2 & G0V & 47.1 $ \pm $ 2.9 & -52.32 $ \pm $ 1.43 & 70.25 $ \pm $ 1.01 \\ 
 44864 & 09:08:27.19 & 27:32:34.62 & 8.3 & G0V & 42.6 $ \pm $ 2.0 & -52.19 $ \pm $ 1.45 & 73.24 $ \pm $ 1.02 \\ 
\hline                                    
 45811 & 09:20:29.03 & -09:33:20.27 & 4.8 & G8III- & 68.2 $ \pm $ 1.4 & -13.42 $ \pm $ 0.90 & -26.80 $ \pm $ 0.57 \\ 
 45802 & 09:20:21.02 & -09:36:36.37 & 7.0 & F4V & 62.8 $ \pm $ 2.7 & -14.57 $ \pm $ 1.21 & -30.04 $ \pm $ 0.78 \\ 
\hline                                    
 45836 & 09:20:43.79 & 51:15:56.57 & 6.1 & F3V & 27.4 $ \pm $ 0.4 & -34.76 $ \pm $ 0.72 & 145.35 $ \pm $ 0.46 \\ 
 45859 & 09:21:03.04 & 51:18:20.41 & 7.8 & G5 & 28.3 $ \pm $ 0.6 & -37.21 $ \pm $ 0.91 & 141.23 $ \pm $ 0.54 \\ 
\hline                                    
 47436 & 09:39:57.85 & 35:20:09.42 & 6.9 & F5 & 49.0 $ \pm $ 1.1 & -67.95 $ \pm $ 0.81 & 2.97 $ \pm $ 0.55 \\ 
 47403 & 09:39:28.10 & 35:14:35.13 & 7.1 & F5 & 49.0 $ \pm $ 1.2 & -68.06 $ \pm $ 0.87 & 2.38 $ \pm $ 0.59 \\ 
\hline                                    
 50325 & 10:16:38.12 & 41:16:33.23 & 7.4 & F5 & 55.8 $ \pm $ 1.8 & 23.71 $ \pm $ 0.73 & -18.18 $ \pm $ 0.67 \\ 
 50327 & 10:16:39.13 & 41:18:19.80 & 8.7 & F8 & 50.1 $ \pm $ 2.3 & 23.27 $ \pm $ 0.92 & -21.44 $ \pm $ 0.78 \\ 
\hline                                    
 52787 & 10:47:31.23 & -22:20:52.67 & 8.4 & K0V & 34.5 $ \pm $ 1.0 & -122.67 $ \pm $ 0.93 & -29.38 $ \pm $ 0.91 \\ 
 52776 & 10:47:25.47 & -22:17:11.91 & 9.9 & K4.5Vk & 32.6 $ \pm $ 1.5 & -125.57 $ \pm $ 1.55 & -30.45 $ \pm $ 1.25 \\ 
\hline                                    
 54692 & 11:11:49.05 & 42:49:57.67 & 7.2 & F8 & 46.8 $ \pm $ 1.3 & -130.55 $ \pm $ 0.65 & -237.88 $ \pm $ 0.66 \\ 
 54681 & 11:11:37.75 & 42:49:05.46 & 8.3 & G5 & 48.4 $ \pm $ 1.9 & -117.50 $ \pm $ 0.85 & -242.43 $ \pm $ 0.81 \\ 
\hline                                    
 57858 & 11:51:57.57 & 08:49:48.42 & 7.4 & K0 & 96.5 $ \pm $ 6.7 & -9.68 $ \pm $ 1.10 & -50.69 $ \pm $ 0.73 \\ 
 57856 & 11:51:56.54 & 08:49:21.97 & 7.9 & K0III- & 98.9 $ \pm $ 7.7 & -1.48 $ \pm $ 1.23 & -52.75 $ \pm $ 0.80 \\ 
\hline                                    
 58067 & 11:54:32.35 & 19:24:40.57 & 8.2 & G0 & 40.5 $ \pm $ 1.8 & -450.28 $ \pm $ 1.40 & -14.27 $ \pm $ 1.01 \\ 
 58073 & 11:54:35.36 & 19:25:40.31 & 8.4 & G5 & 38.4 $ \pm $ 2.0 & -452.68 $ \pm $ 1.32 & -15.39 $ \pm $ 1.15 \\ 
\hline                                    
 58751 & 12:02:59.82 & -10:45:08.45 & 7.4 & F5 & 59.2 $ \pm $ 2.2 & 33.82 $ \pm $ 0.73 & -18.37 $ \pm $ 0.60 \\ 
 58722 & 12:02:39.44 & -10:42:48.94 & 8.5 & G0 & 55.1 $ \pm $ 3.0 & 33.59 $ \pm $ 0.92 & -17.81 $ \pm $ 0.72 \\ 
\hline                                    
 64057 & 13:07:39.88 & 24:00:34.03 & 8.2 & G5V & 35.4 $ \pm $ 1.2 & -260.82 $ \pm $ 1.07 & 148.28 $ \pm $ 0.81 \\ 
 64059 & 13:07:40.90 & 24:01:10.47 & 8.6 & G0 & 37.3 $ \pm $ 1.7 & -261.79 $ \pm $ 1.15 & 146.19 $ \pm $ 0.92 \\ 
\hline                                    
 64444 & 13:12:32.07 & -34:44:50.60 & 7.8 & F5/F6V & 77.0 $ \pm $ 6.9 & -250.44 $ \pm $ 1.28 & -270.47 $ \pm $ 1.56 \\ 
 64443 & 13:12:30.40 & -34:45:14.27 & 9.8 & K0 & 58.8 $ \pm $ 7.5 & -244.24 $ \pm $ 2.99 & -274.78 $ \pm $ 2.46 \\ 
\hline                                    
 65602 & 13:27:03.12 & -24:17:25.06 & 8.7 & K2+v & 29.4 $ \pm $ 0.7 & -337.25 $ \pm $ 1.34 & -61.08 $ \pm $ 0.70 \\ 
 65574 & 13:26:39.77 & -24:17:35.52 & 8.8 & K2.5Vk & 30.2 $ \pm $ 1.0 & -335.87 $ \pm $ 1.57 & -70.65 $ \pm $ 0.83 \\ 
\hline                                    
 65899 & 13:30:30.77 & 22:30:47.47 & 8.6 & F8 & 76.2 $ \pm $ 5.2 & 27.20 $ \pm $ 1.08 & -45.48 $ \pm $ 0.65 \\ 
 65884 & 13:30:18.97 & 21:30:01.38 & 10.0 & G5 & 71.9 $ \pm $ 6.4 & 20.33 $ \pm $ 1.55 & -49.68 $ \pm $ 0.85 \\ 
\hline                                    
 66749 & 13:40:50.75 & -17:47:33.53 & 7.9 & F3V & 85.1 $ \pm $ 4.7 & -35.86 $ \pm $ 1.05 & -11.49 $ \pm $ 0.66 \\ 
 66717 & 13:40:32.53 & -17:20:37.30 & 8.5 & F3V & 85.7 $ \pm $ 6.2 & -46.34 $ \pm $ 1.39 & -13.34 $ \pm $ 0.65 \\ 
\hline                                    
 67250 & 13:46:59.87 & 38:32:33.90 & 5.5 & K0III+ & 97.2 $ \pm $ 3.1 & -134.18 $ \pm $ 0.47 & -21.69 $ \pm $ 0.45 \\ 
 67041 & 13:44:20.40 & 38:47:52.05 & 8.9 & G5 & 95.2 $ \pm $ 9.6 & -141.29 $ \pm $ 0.72 & -21.72 $ \pm $ 0.75 \\ 
\hline                                    
 68830 & 14:05:38.54 & -18:04:20.08 & 8.2 & F8+... & 80.6 $ \pm $ 7.2 & 41.70 $ \pm $ 1.41 & -44.51 $ \pm $ 1.08 \\ 
 68833 & 14:05:38.92 & -18:04:50.92 & 8.8 & G0 & 78.2 $ \pm $ 7.3 & 37.18 $ \pm $ 1.80 & -44.64 $ \pm $ 1.59 \\ 
\hline                                    
 69701 & 14:16:00.88 & -05:59:58.29 & 4.1 & F7IV & 22.2 $ \pm $ 0.1 & -25.84 $ \pm $ 0.91 & -419.84 $ \pm $ 0.68 \\ 
 69962 & 14:18:58.28 & -06:36:09.34 & 9.1 & K7V & 21.8 $ \pm $ 0.8 & -5.17 $ \pm $ 1.89 & -432.05 $ \pm $ 1.29 \\ 
\hline                                    
 71726 & 14:40:18.33 & 30:26:38.20 & 7.7 & G0 & 53.5 $ \pm $ 1.7 & 95.06 $ \pm $ 0.89 & -44.25 $ \pm $ 0.77 \\ 
 71737 & 14:40:28.34 & 30:31:14.23 & 8.0 & G2IV & 51.3 $ \pm $ 2.4 & 95.59 $ \pm $ 1.06 & -44.78 $ \pm $ 0.86 \\ 
\hline                                    
 74442 & 15:12:47.73 & 27:55:36.72 & 8.4 & G0V & 79.8 $ \pm $ 5.5 & 9.63 $ \pm $ 1.10 & -131.12 $ \pm $ 0.93 \\ 
 74439 & 15:12:45.92 & 27:55:15.21 & 9.4 & G0 & 69.1 $ \pm $ 6.4 & 7.13 $ \pm $ 1.69 & -131.25 $ \pm $ 1.41 \\ 
\hline                                    
 74666 & 15:15:30.10 & 33:18:54.37 & 3.5 & G8III & 37.3 $ \pm $ 0.2 & 84.84 $ \pm $ 0.37 & -110.57 $ \pm $ 0.48 \\ 
 74674 & 15:15:38.29 & 33:19:16.28 & 7.8 & G0Vv & 37.0 $ \pm $ 1.0 & 83.89 $ \pm $ 0.52 & -109.54 $ \pm $ 0.68 \\ 
\hline                                    
 85620 & 17:29:44.45 & 63:51:11.15 & 7.7 & F9V & 45.2 $ \pm $ 1.0 & 2.48 $ \pm $ 0.63 & -182.16 $ \pm $ 0.67 \\ 
 85575 & 17:29:16.58 & 63:52:10.49 & 8.4 & G0 & 45.2 $ \pm $ 1.1 & 0.31 $ \pm $ 0.72 & -181.38 $ \pm $ 0.77 \\ 
\hline                                    
 96895 & 19:41:49.09 & 50:31:31.61 & 6.0 & G1.5Vb & 21.1 $ \pm $ 0.1 & -147.75 $ \pm $ 0.56 & -158.85 $ \pm $ 0.50 \\ 
 96901 & 19:41:52.10 & 50:31:04.51 & 6.2 & G3V & 21.2 $ \pm $ 0.1 & -135.15 $ \pm $ 0.60 & -163.53 $ \pm $ 0.52 \\ 
\hline                                    
 97295 & 19:46:25.58 & 33:43:43.28 & 5.0 & F7V & 21.2 $ \pm $ 0.1 & 23.05 $ \pm $ 0.42 & -448.66 $ \pm $ 0.50 \\ 
 97222 & 19:45:33.52 & 33:36:11.00 & 7.7 & K3V & 21.1 $ \pm $ 0.4 & 13.30 $ \pm $ 1.07 & -440.57 $ \pm $ 1.35 \\ 
\hline                                    
 99729 & 20:14:09.85 & 06:35:20.83 & 7.7 & G4IV & 61.1 $ \pm $ 4.6 & -132.91 $ \pm $ 1.57 & -60.63 $ \pm $ 1.84 \\ 
 99727 & 20:14:09.26 & 06:34:38.49 & 8.0 & G4V & 61.4 $ \pm $ 5.2 & -130.02 $ \pm $ 1.74 & -61.72 $ \pm $ 2.06 \\ 
\hline                                    
 101082 & 20:29:27.31 & 81:05:26.66 & 6.0 & K0III+ & 63.7 $ \pm $ 0.9 & 67.05 $ \pm $ 0.62 & 221.67 $ \pm $ 0.55 \\ 
 101166 & 20:30:21.65 & 81:08:20.21 & 8.7 & G5 & 68.9 $ \pm $ 2.4 & 67.54 $ \pm $ 0.87 & 221.35 $ \pm $ 0.81 \\ 
\hline                                    
 101916 & 20:39:07.59 & 10:05:10.15 & 5.1 & G5IV+. & 30.1 $ \pm $ 0.3 & 323.58 $ \pm $ 0.82 & 21.07 $ \pm $ 0.55 \\ 
 101932 & 20:39:21.86 & 10:04:32.46 & 8.5 & K2V & 28.7 $ \pm $ 0.7 & 317.09 $ \pm $ 1.27 & 20.21 $ \pm $ 0.84 \\ 
\hline                                    
 110419 & 22:21:57.60 & -34:31:12.96 & 7.4 & F4V & 86.2 $ \pm $ 7.7 & 9.87 $ \pm $ 1.21 & 25.83 $ \pm $ 0.85 \\ 
 110433 & 22:22:04.58 & -34:29:20.30 & 7.5 & F2/F3I & 74.0 $ \pm $ 4.2 & 9.65 $ \pm $ 1.13 & 26.62 $ \pm $ 0.68 \\ 
\hline                                    
 112222 & 22:43:42.69 & 10:56:23.23 & 6.5 & G8IV & 42.4 $ \pm $ 3.0 & 9.80 $ \pm $ 1.11 & -171.17 $ \pm $ 1.25 \\ 
 112354 & 22:45:27.86 & 11:11:32.43 & 9.8 & K6V: & 41.6 $ \pm $ 2.8 & 13.82 $ \pm $ 2.96 & -162.90 $ \pm $ 2.92 \\ 
\hline                                    
 112970 & 22:52:42.16 & 67:59:24.04 & 7.0 & F2 & 59.0 $ \pm $ 1.5 & 78.49 $ \pm $ 0.67 & 63.42 $ \pm $ 0.61 \\ 
 112946 & 22:52:30.05 & 67:59:36.49 & 7.5 & F5 & 59.9 $ \pm $ 1.4 & 80.88 $ \pm $ 0.56 & 59.22 $ \pm $ 0.52 \\ 
\hline                                    
 113579 & 23:00:19.22 & -26:09:12.09 & 7.5 & G5V & 30.8 $ \pm $ 0.7 & 108.87 $ \pm $ 1.02 & -160.41 $ \pm $ 0.74 \\ 
 113597 & 23:00:27.88 & -26:18:41.38 & 9.6 & K7V & 30.5 $ \pm $ 1.9 & 113.63 $ \pm $ 2.65 & -162.16 $ \pm $ 2.11 \\ 
\hline                                    
 117573 & 23:50:37.98 & 54:11:53.91 & 7.1 & F5 & 58.5 $ \pm $ 2.2 & -2.18 $ \pm $ 0.54 & -34.03 $ \pm $ 0.55 \\ 
 117733 & 23:52:39.67 & 54:16:08.14 & 7.6 & F5 & 55.9 $ \pm $ 2.0 & -5.13 $ \pm $ 0.64 & -39.66 $ \pm $ 0.67 \\ 
\hline                                    
 118254 & 23:59:08.97 & 41:12:06.25 & 7.7 & G0 & 44.3 $ \pm $ 1.3 & 80.21 $ \pm $ 0.68 & 4.40 $ \pm $ 0.51 \\ 
 118251 & 23:59:06.80 & 41:10:13.97 & 8.2 & G0 & 41.9 $ \pm $ 1.1 & 82.20 $ \pm $ 0.70 & 3.56 $ \pm $ 0.52 \\ 
\end{longtable}

\section{Observations and data reduction}

Observations were gathered during two separated observing runs, in
January and December 2013, at the 2.5 m Nordic Optical Telescope,
located at the Roque de los Muchachos observatory, in the Canary
islands. Radial velocities were obtained with the fiber feed FIES
Echelle spectrograph (Ref.~\refcite{telting14}), in high resolution mode
($R = \lambda / \Delta \lambda = 67000$) plus simultaneous 
calibration lamp to achieve the highest
velocity accuracy.  Atmospheric conditions were below average (poor
seeing) during observations, nevertheless targets were bright enough
to ensure spectra with sufficiently high signal to noise spectra for our purpose
could be gathered within minutes.  Note that to ensure the best stability 
the FIES spectrograph is
sitting in a dedicated protected room, and is linked to the telescope
by an optical fiber. Moreover, in all cases, the two components of a
given pair were observed consecutively one ofter the other within minutes,
ensuring environmental variation (e.g., pressure and temperature) had no effect.

Data have been reduced using the FIES pipeline (Ref.~\refcite{stemples05}).
The pipeline performs the standard reduction, bias subtraction, flat
fielding, identification of the spectra position and extraction,
combination of the different orders, wavelength calibration using a
separate calibration frame and, finally, it uses the footprint of the simultaneous
calibration lamp to compensate for any wavelength instability in the
spectrograph.

\section{Spectral analysis}

\begin{figure*}[htbp]
\centering
\includegraphics[width=0.49\columnwidth]{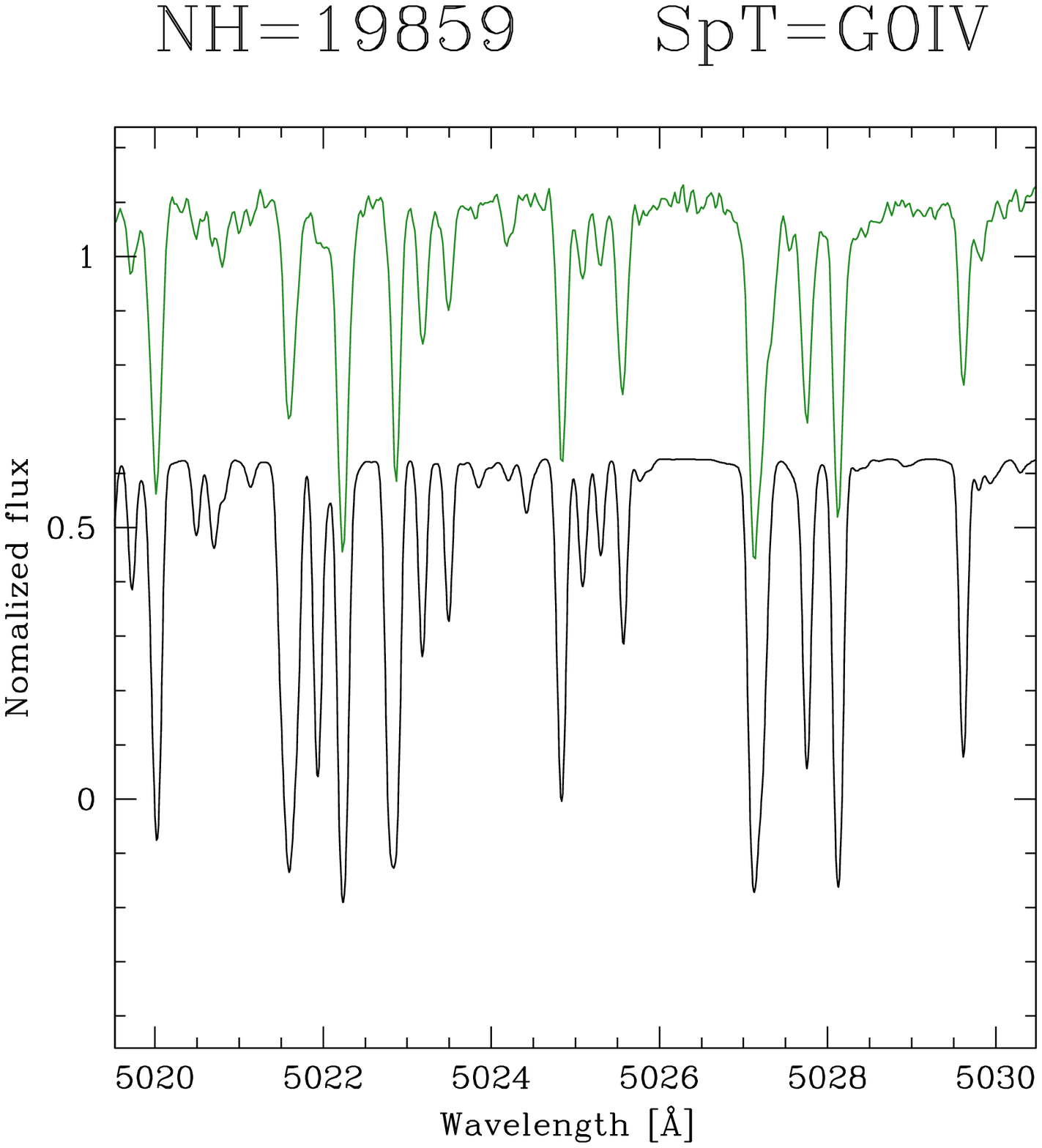}
\includegraphics[width=0.49\columnwidth]{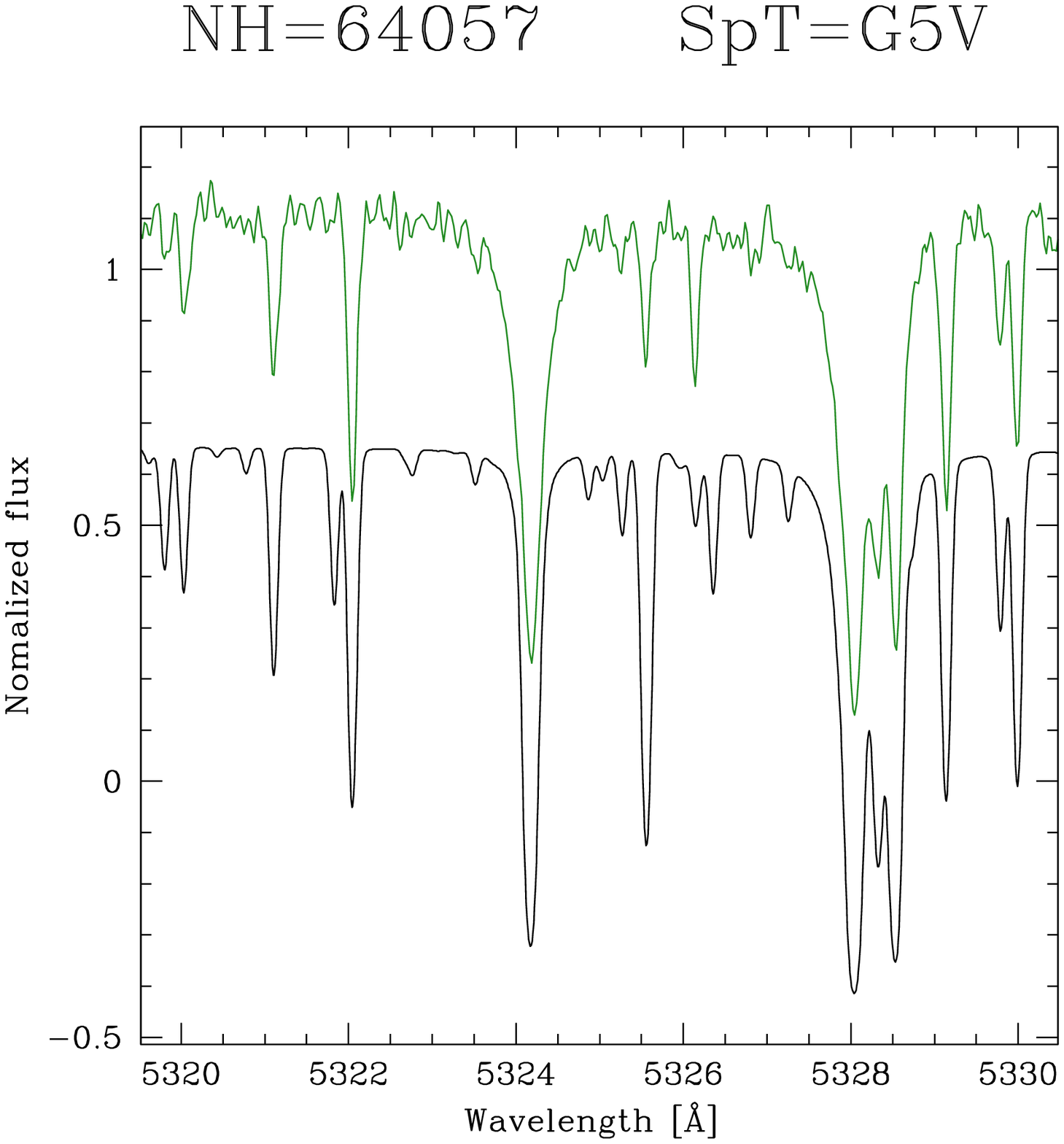}
\includegraphics[width=0.49\columnwidth]{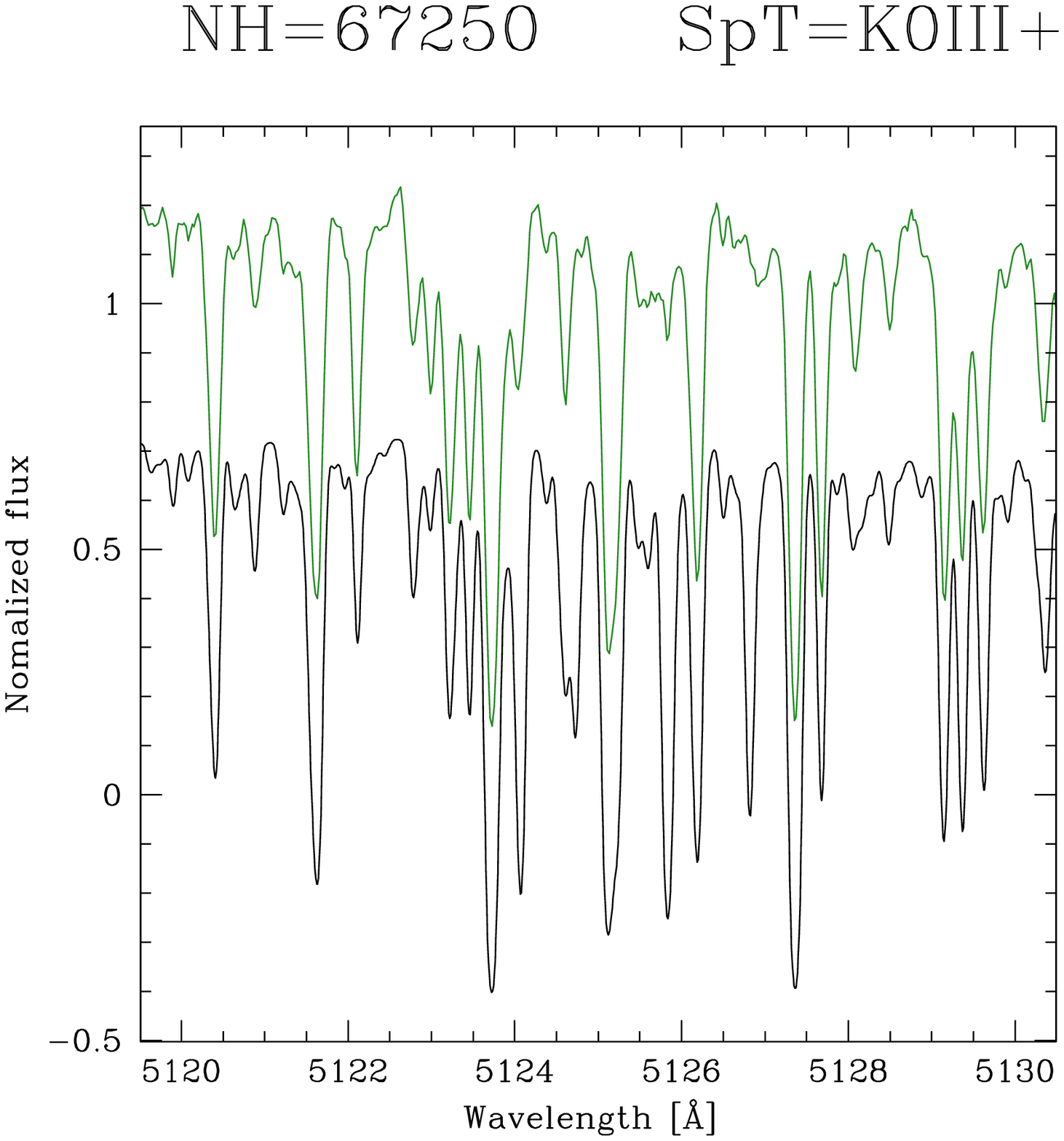}
\includegraphics[width=0.49\columnwidth]{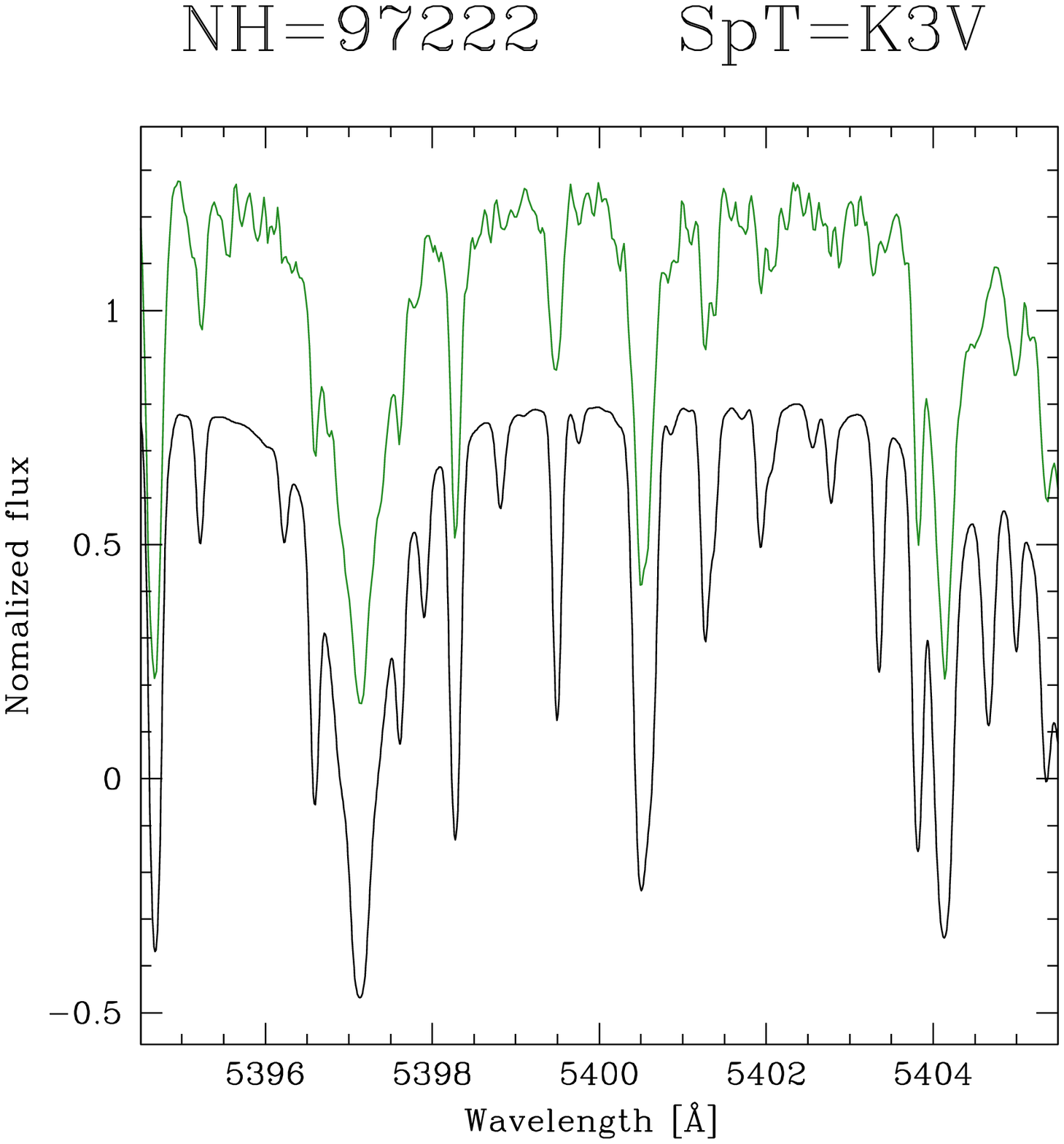}
\caption{Examples of observed spectrum (green top spectrum) 
with the adopted stellar synthetic template (black bottom spectrum) for stars 
of various spectral type. }
\label{fig:sptempl}
\end{figure*}

The crucial aspect of our analysis is the measurement of the radial
velocity of the selected stars.  To do this we performed
cross-correlation (task FXCOR in IRAF) of the observed stellar
spectrum with an appropriate template.  Synthetic template spectra
were extracted from the BLUERED
library\footnote[1]{http://inaoep.mx/~modelos/bluered/bluered.html},
an extended theoretical library of synthetic stellar
spectra (Ref.~\refcite{bertone08}), covering the optical range from
3500 to 7000 Å at a spectral resolving power $R = 500000$. The library
is based on the ATLAS9 model atmospheres (Ref.~\refcite{kurucz95}).
The grid spans a large volume in three-dimensional parameters space:
the effective temperature ranges from 4000 to 50,000 K and the surface
gravity spans from 0.0 to 5.0 dex at six different solar-scaled
metallicity values ([M/H] = -3.0, -2.0, -1.0, -0.3, +0.0, +0.3).
Template-spectra were degraded to R=70000 to match the resolution of
the data.  Examples of observed spectra and the associated templates
are shown in fig.  \ref{fig:sptempl}.  Metallicity has marginal
influence in the templates, thus we assume solar metallicity for all
spectra.  The association of the template to each observed spectrum
was based on the spectral type of the star and we checked the
similarity by visual inspection.\\

\begin{figure*}[htbp]
\centering

\includegraphics[width=0.49\columnwidth]{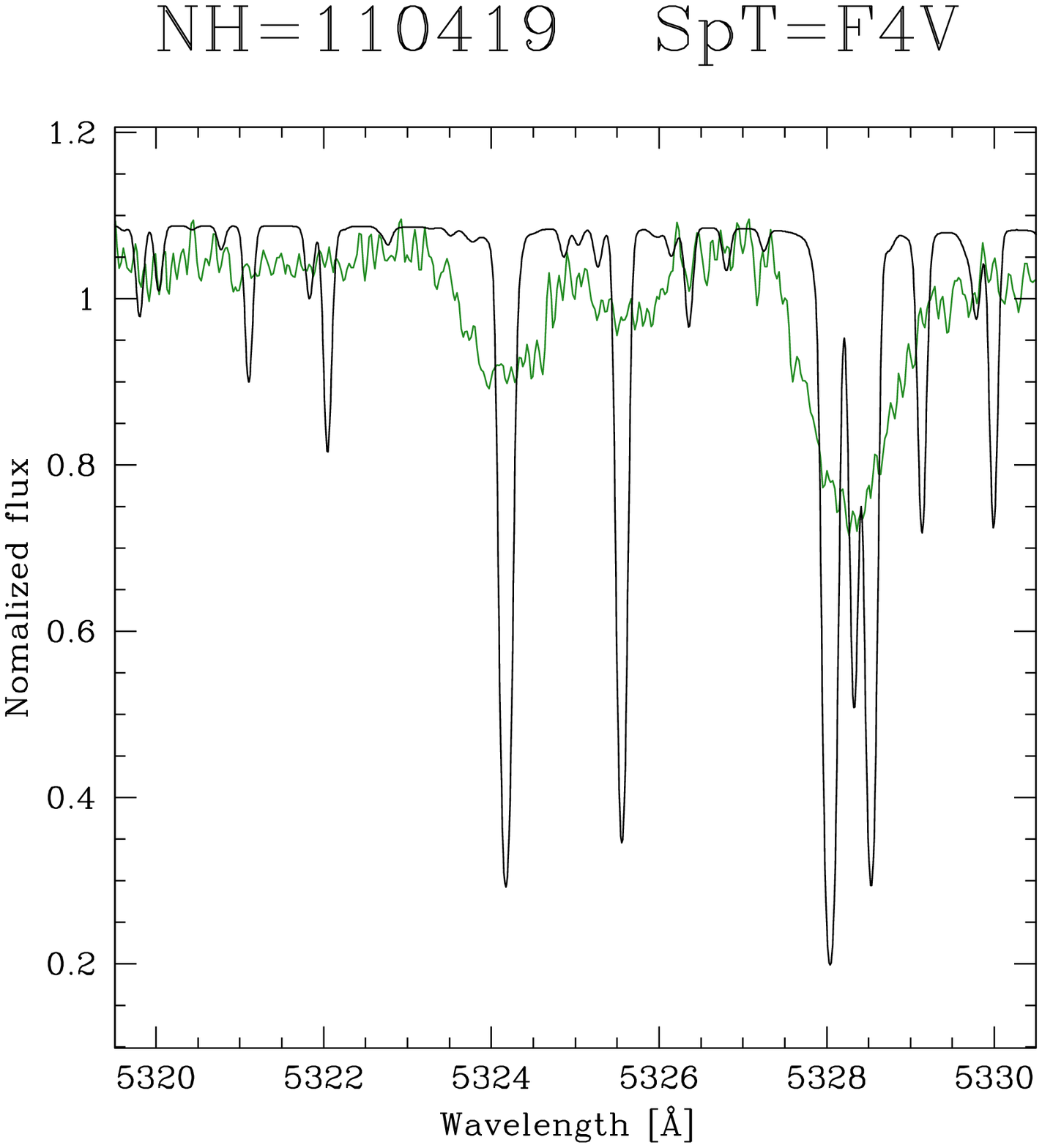}
\includegraphics[width=0.49\columnwidth]{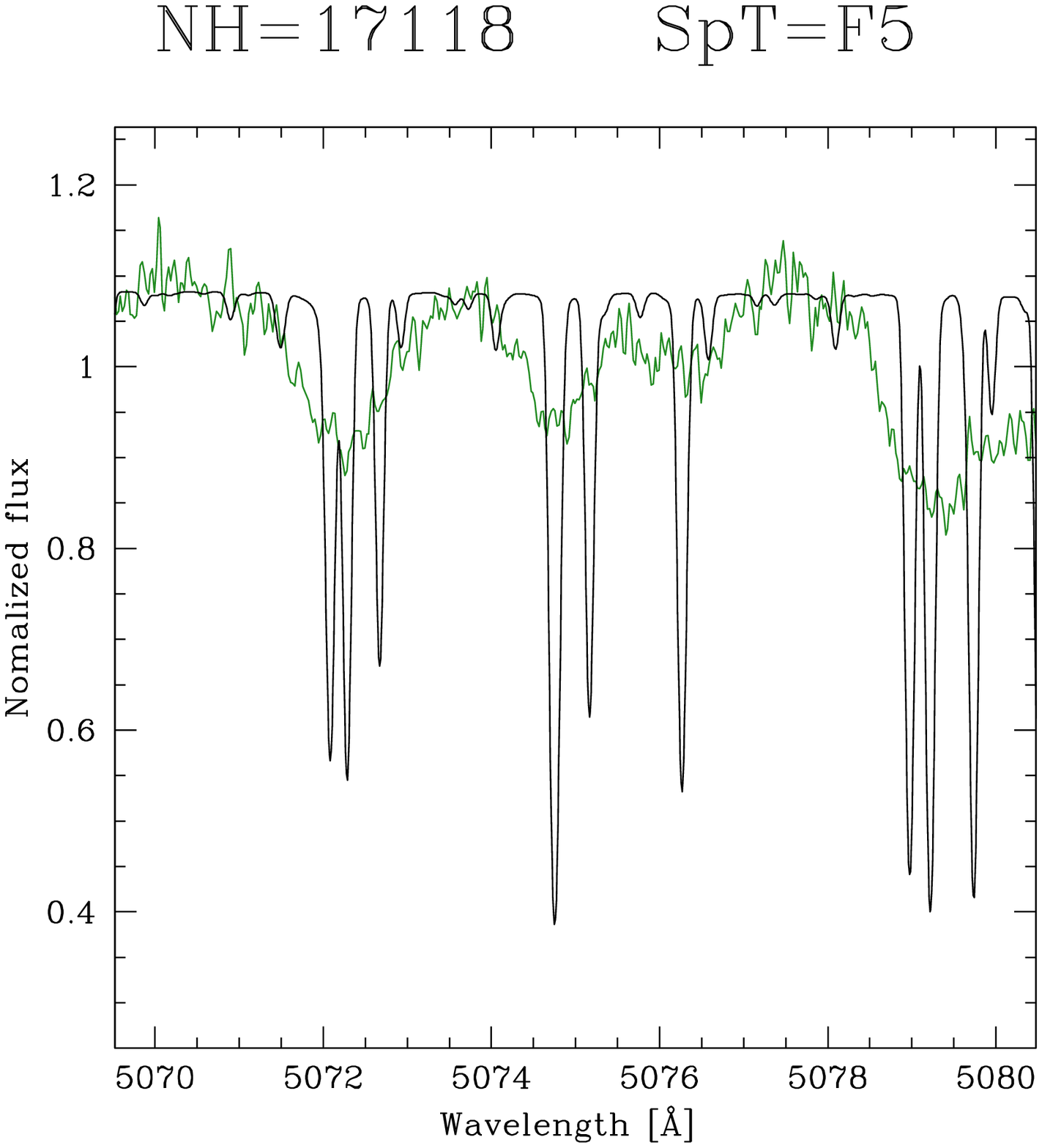}

\includegraphics[width=0.49\columnwidth]{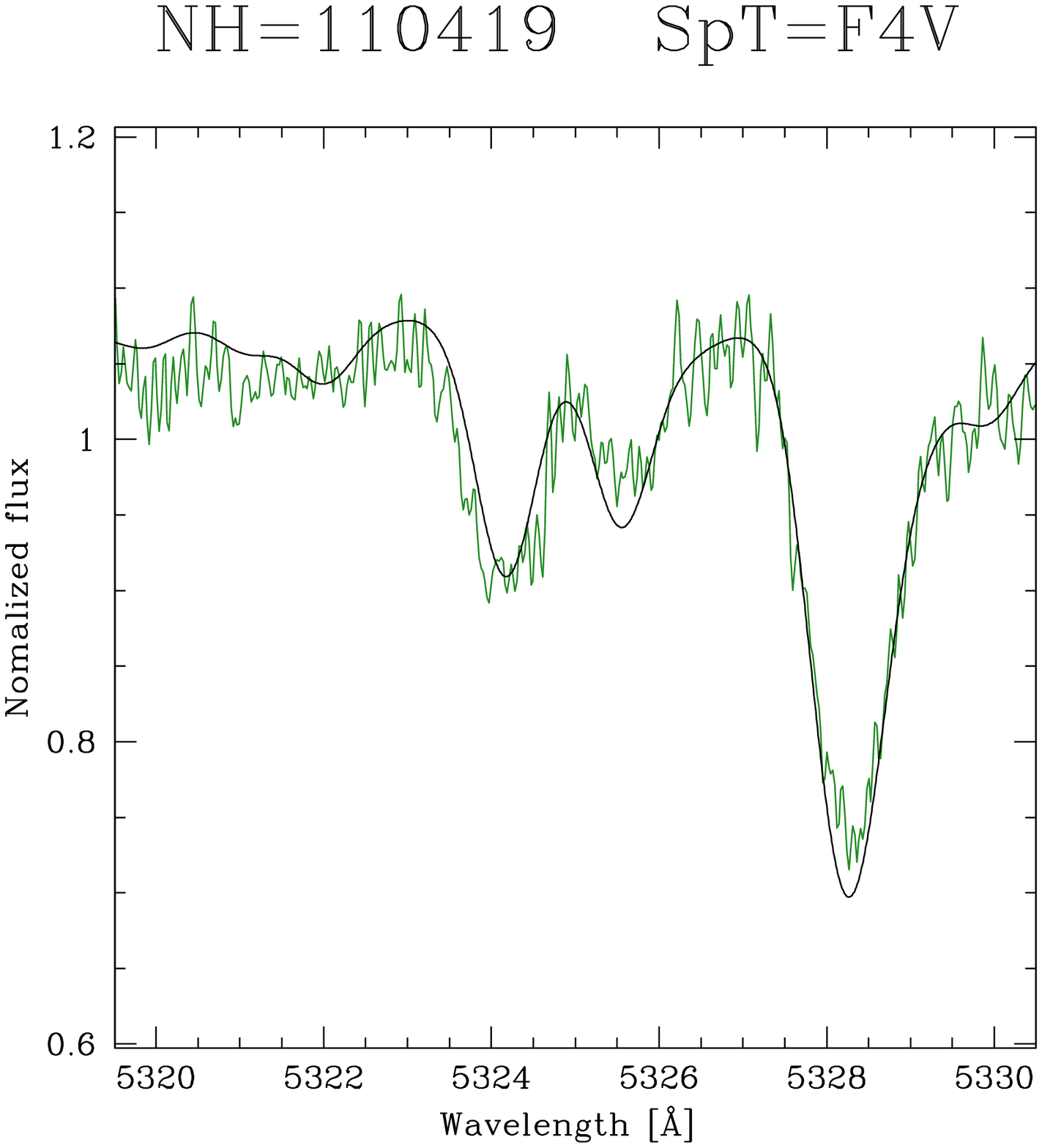}
\includegraphics[width=0.49\columnwidth]{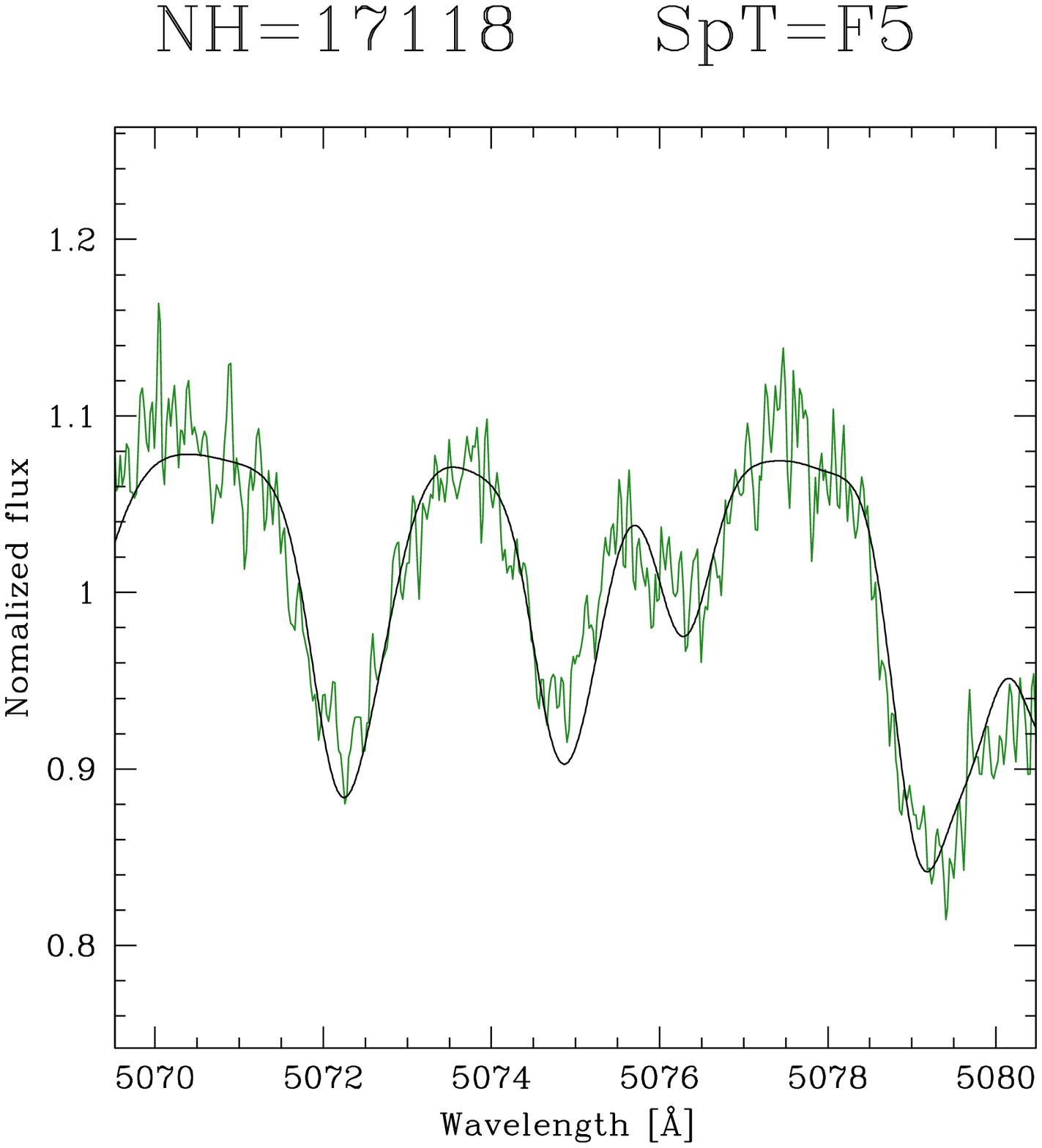}

\caption{Upper panels: Example of observed spectrum (green) compared to the best associated 
template (black) before convolution to compensate for the star rotation. 
The difference in line width
is apparent. Lower panels: Spectra and template after convolution to match
line width. 
 }
\label{fig:spconv}
\end{figure*}

\indent In a number of cases the observed spectra exhibit absorption
lines that are much wider than expected for the corresponding spectral
type.  This broadening of the lines is interpreted as due to fast
rotation of the star.  An effect most evident in stars of spectral
type F5 or earlier. To cope with this problem templates were convolved
with a Gaussian kernels of various sizes to match the observed line
width (Fig. \ref{fig:spconv}). The final cross-correlation with the
modified template spectrum did work well, though the measurements of
the radial velocity have somewhat larger errors in these cases.

\subsection{Analysis}

For each spectrum, cross-correlation was performed in spectral
intervals of 200 \mbox{\AA} between 4100 \mbox{\AA} and 6700
\mbox{\AA} to derive the radial velocity (RV) of the stars. Excluding
the spectral regions contaminated by telluric absorption bands, there
are 13 suitable intervals, resulting in 13 values of RV.
\ \\ \ \\ \indent Then, for each interval, the measured RV was
retained according to the following criteria:
\begin{itemlist}
 \item correlation peak greater than 0.5;
 \item error on RV smaller than 2$\sigma$ from the average error (of the other intervals);
 \item RV within  5$\sigma$ from average of RV ($\sigma\rm{-clipping}$).
\end{itemlist}

In all cases at least 8 out of 13 values could be retained. The final
RV associated to each star was then computed as the weighted mean of
these values.  Tests were made to check whether our measurements could
be sensitive to small changes in the selected template. No such effect
was found.  The final velocities, corrected to heliocentric reference
frame, are given in Table \ref{Tab:2}.  

\begin{center}
\begin{longtable}{cccc}

\caption{Table of results (horizontal lines separate different pairs).\\
Legend:\\
\textbf{NH} = Hipparcos catalogue star number;\\
\textbf{MJD} = Modified julian date of observation;\\
\textbf{expt} = Exposure time in seconds;\\
\textbf{RV} = Measured heliocentric radial velocity in km/s.\\
} \label{Tab:2}\\

\hline
NH & MJD & expt & RV \\
 & [day] & [s]  & [km/s]\\                     
\hline
\endfirsthead

\multicolumn{4}{l}{{\bfseries \tablename\ \thetable{} -- continued from previous page}} \\
\hline
NH & MJD & expt & RV \\
 & [day] & [s]  & [km/s]\\                   
\hline
\endhead

\hline
\multicolumn{4}{r}{ \bfseries Continued on next page}
\endfoot

\hline
\endlastfoot
 185    & 56643.924 & 350 &$   0.830  \pm  0.016$ \\ 
 190    & 56643.931 & 400 &$   0.505  \pm  0.014$ \\ 
\hline                                           
 201    & 56644.877 & 600 &$   3.969  \pm  0.058$ \\ 
 206    & 56644.888 & 700 &$   3.659  \pm  0.080$ \\ 
\hline                                           
 1891   & 56324.829 & 200 &$ -24.515  \pm  0.019$ \\ 
 1887   & 56324.836 & 300 &$ -20.508  \pm  0.015$ \\ 
\hline                                           
 2292   & 56324.844 & 200 &$  10.105  \pm  0.014$ \\ 
 2350   & 56324.851 & 600 &$   9.854  \pm  0.018$ \\ 
\hline                                           
 4702   & 56644.931 & 400 &$   5.647  \pm  0.012$ \\ 
 4833   & 56644.940 & 450 &$   6.103  \pm  0.014$ \\ 
\hline                                           
 8497   & 56643.959 & 120 &$  -1.814  \pm  0.250$ \\ 
 8486   & 56643.964 & 220 &$  -4.905  \pm  0.013$ \\ 
\hline                                           
 10321  & 56324.881 & 200 &$   7.252  \pm  0.014$ \\ 
 10339  & 56324.887 & 200 &$   7.037  \pm  0.018$ \\ 
\hline                                           
 11137  & 56643.987 & 500 &$  27.129  \pm  0.014$ \\ 
 11134  & 56643.996 & 600 &$  27.286  \pm  0.016$ \\ 
\hline                                           
 11783  & 56644.008 &  90 &$ -26.254  \pm  0.018$ \\ 
 11759  & 56644.013 & 500 &$ -27.912  \pm  0.021$ \\ 
\hline                                           
 15304  & 56324.915 & 250 &$  31.860  \pm  0.019$ \\ 
 15310  & 56324.921 & 300 &$  32.716  \pm  0.019$ \\ 
\hline                                           
 15527  & 56644.950 & 350 &$  40.287  \pm  0.019$ \\ 
 15526  & 56644.960 & 550 &$  40.664  \pm  0.014$ \\ 
\hline                                           
 17118  & 56324.930 & 200 &$  17.006  \pm  0.177$ \\ 
 17126  & 56324.936 & 400 &$  12.288  \pm  0.043$ \\ 
\hline                                           
 19335  & 56323.825 &  60 &$  26.461  \pm  0.054$ \\ 
 19335  & 56324.945 &  70 &$  26.498  \pm  0.055$ \\ 
 19255  & 56323.834 & 120 &$  27.846  \pm  0.016$ \\ 
 19255  & 56324.950 & 250 &$  28.030  \pm  0.017$ \\ 
\hline                                           
 19859  & 56323.842 &  90 &$  -6.858  \pm  0.012$ \\ 
 19855  & 56323.848 &  90 &$  -7.430  \pm  0.016$ \\ 
\hline                                           
 20342  & 56324.961 & 500 &$  78.756  \pm  0.016$ \\ 
 20338  & 56645.020 & 1200&$  89.218  \pm  0.057$ \\ 
\hline                                           
 20669  & 56323.854 & 150 &$ -35.784  \pm  0.026$ \\ 
 20637  & 56323.864 & 600 &$ -36.285  \pm  0.017$ \\ 
\hline                                           
 21537  & 56323.913 & 150 &$  38.947  \pm  0.016$ \\ 
 21534  & 56323.919 & 150 &$  39.082  \pm  0.015$ \\ 
\hline                                           
 21704  & 56644.064 & 400 &$   8.374  \pm  0.026$ \\ 
 21702  & 56644.073 & 600 &$   8.417  \pm  0.015$ \\ 
\hline                                           
 22611  & 56645.056 & 250 &$  46.203  \pm  0.015$ \\ 
 22604  & 56645.063 & 500 &$  46.444  \pm  0.016$ \\ 
\hline                                           
 24046  & 56323.925 & 120 &$  15.889  \pm  0.022$ \\ 
 24046  & 56324.977 & 300 &$  16.001  \pm  0.024$ \\ 
 24035  & 56323.930 & 480 &$   7.666  \pm  0.023$ \\ 
 24035  & 56324.989 & 900 &$   7.775  \pm  0.023$ \\ 
\hline                                           
 25278  & 56323.941 &  45 &$  38.350  \pm  0.047$ \\ 
 25220  & 56323.945 & 200 &$  38.603  \pm  0.023$ \\ 
\hline                                           
 33705  & 56644.130 & 150 &$  16.700  \pm  0.027$ \\ 
 33691  & 56644.139 & 600 &$  17.141  \pm  0.017$ \\ 
\hline                                           
 34426  & 56325.042 & 300 &$ -11.552  \pm  0.019$ \\ 
 34426  & 56644.150 & 400 &$ -11.538  \pm  0.017$ \\ 
 34407  & 56325.050 & 300 &$ -12.309  \pm  0.018$ \\ 
 34407  & 56644.163 & 400 &$ -12.350  \pm  0.014$ \\ 
\hline                                           
 34714  & 56325.057 & 350 &$   3.196  \pm  0.020$ \\ 
 34714  & 56646.121 & 900 &$   3.196  \pm  0.018$ \\ 
 34700  & 56325.064 & 500 &$   3.490  \pm  0.021$ \\ 
 34700  & 56646.139 & 1200&$   3.364  \pm  0.017$ \\ 
\hline                                           
 39457  & 56645.111 & 500 &$  26.807  \pm  0.015$ \\ 
 39452  & 56645.121 & 600 &$  27.548  \pm  0.016$ \\ 
\hline                                           
 40918  & 56323.978 & 200 &$   8.402  \pm  0.017$ \\ 
 40918  & 56645.147 & 500 &$   5.953  \pm  0.014$ \\ 
 40882  & 56323.986 & 300 &$   2.741  \pm  0.018$ \\ 
 40882  & 56645.157 & 550 &$   2.749  \pm  0.015$ \\ 
\hline                                           
 44520  & 56325.115 & 600 &$   5.461  \pm  0.015$ \\ 
 44520  & 56646.176 & 900 &$   6.626  \pm  0.016$ \\ 
 44518  & 56325.128 & 700 &$   7.474  \pm  0.016$ \\ 
 44518  & 56646.190 & 1000&$   6.461  \pm  0.018$ \\ 
\hline                                           
 44858  & 56324.021 & 250 &$  30.478  \pm  0.015$ \\ 
 44858  & 56645.214 & 500 &$  30.469  \pm  0.014$ \\ 
 44864  & 56324.028 & 250 &$  30.746  \pm  0.017$ \\ 
 44864  & 56645.223 & 500 &$  30.781  \pm  0.015$ \\ 
\hline                                           
 45811  & 56325.140 &  60 &$  25.246  \pm  0.014$ \\ 
 45811  & 56644.241 & 200 &$  25.312  \pm  0.013$ \\ 
 45802  & 56325.144 & 150 &$  36.886  \pm  0.091$ \\ 
 45802  & 56644.247 & 400 &$  -2.116  \pm  0.071$ \\ 
\hline                                           
 45836  & 56324.035 &  90 &$  -7.789  \pm  0.016$ \\ 
 45836  & 56645.200 & 120 &$  -7.730  \pm  0.017$ \\ 
 45859  & 56324.039 & 200 &$  -6.436  \pm  0.014$ \\ 
 45859  & 56645.205 & 400 &$  -6.414  \pm  0.013$ \\ 
\hline                                           
 47436  & 56325.180 & 120 &$   6.288  \pm  0.031$ \\ 
 47436  & 56646.229 & 300 &$   6.272  \pm  0.030$ \\ 
 47403  & 56325.185 & 300 &$ -34.218  \pm  0.017$ \\ 
 47403  & 56646.236 & 600 &$ -36.193  \pm  0.018$ \\ 
\hline                                           
 50325  & 56324.046 & 150 &$  10.812  \pm  0.032$ \\ 
 50327  & 56324.055 & 300 &$  12.289  \pm  0.015$ \\ 
\hline                                           
 52787  & 56325.219 & 250 &$  24.212  \pm  0.019$ \\ 
 52776  & 56325.225 & 500 &$  24.622  \pm  0.022$ \\ 
\hline                                           
 54692  & 56324.081 & 150 &$  11.727  \pm  0.015$ \\ 
 54692  & 56645.233 & 400 &$  11.665  \pm  0.015$ \\ 
 54681  & 56324.088 & 250 &$  10.346  \pm  0.014$ \\ 
 54681  & 56645.242 & 600 &$  10.341  \pm  0.013$ \\ 
\hline                                           
 57858  & 56325.234 & 120 &$  -4.031  \pm  0.013$ \\ 
 57856  & 56325.238 & 150 &$  -3.601  \pm  0.048$ \\ 
\hline                                           
 58067  & 56325.243 & 200 &$   6.461  \pm  0.015$ \\ 
 58073  & 56325.248 & 200 &$   6.320  \pm  0.014$ \\ 
\hline                                           
 58751  & 56324.101 & 250 &$ -12.248  \pm  0.041$ \\ 
 58722  & 56324.107 & 300 &$ -15.411  \pm  0.020$ \\ 
\hline                                           
 64057  & 56324.116 & 300 &$  -1.247  \pm  0.014$ \\ 
 64059  & 56324.123 & 500 &$  -1.264  \pm  0.013$ \\ 
\hline                                           
 64444  & 56325.254 & 120 &$  -8.575  \pm  0.027$ \\ 
 64443  & 56325.258 & 500 &$  -3.156  \pm  0.021$ \\ 
\hline                                           
 65602  & 56644.285 & 400 &$ -12.884  \pm  0.018$ \\ 
 65574  & 56644.294 & 450 &$ -11.870  \pm  0.018$ \\ 
\hline                                           
 65899  & 56324.134 & 500 &$ -35.396  \pm  0.019$ \\ 
 65884  & 56324.144 & 900 &$  57.917  \pm  0.015$ \\ 
\hline                                           
 66749  & 56324.248 & 300 &$  -7.355  \pm  0.195$ \\ 
 66749  & 56325.277 & 120 &$  -7.909  \pm  0.055$ \\ 
 66717  & 56324.255 & 400 &$  -0.809  \pm  0.031$ \\ 
 66717  & 56325.282 & 240 &$  -0.913  \pm  0.036$ \\ 
\hline                                           
 67250  & 56324.233 &  60 &$ -10.781  \pm  0.014$ \\ 
 67250  & 56325.296 &  60 &$ -10.776  \pm  0.014$ \\ 
 67041  & 56324.237 & 500 &$ -32.428  \pm  0.012$ \\ 
 67041  & 56325.300 & 150 &$ -32.393  \pm  0.012$ \\ 
\hline                                           
 68830  & 56324.203 & 300 &$  22.022  \pm  0.280$ \\ 
 68833  & 56324.210 & 480 &$  28.301  \pm  0.016$ \\ 
\hline                                           
 69701  & 56324.264 &  60 &$  12.155  \pm  0.041$ \\ 
 69962  & 56324.269 & 600 &$  11.817  \pm  0.041$ \\ 
\hline                                           
 71726  & 56324.281 & 200 &$ -11.468  \pm  0.015$ \\ 
 71737  & 56324.288 & 480 &$ -11.298  \pm  0.015$ \\ 
\hline                                           
 74442  & 56646.248 & 600 &$ -60.207  \pm  0.016$ \\ 
 74439  & 56646.259 & 900 &$ -59.245  \pm  0.016$ \\ 
\hline                                           
 74666  & 56646.273 &  90 &$ -11.844  \pm  0.013$ \\ 
 74674  & 56646.277 & 700 &$ -11.477  \pm  0.017$ \\ 
\hline                                           
 85620  & 56644.815 & 350 &$ -33.736  \pm  0.027$ \\ 
 85575  & 56644.824 & 450 &$ -33.294  \pm  0.021$ \\ 
\hline                                           
 96895  & 56644.800 & 200 &$ -27.455  \pm  0.011$ \\ 
 96901  & 56644.809 & 200 &$ -27.043  \pm  0.011$ \\ 
\hline                                           
 97295  & 56644.787 &  90 &$   4.769  \pm  0.023$ \\ 
 97222  & 56644.794 & 200 &$   5.124  \pm  0.015$ \\ 
\hline                                           
 99729  & 56644.833 & 450 &$  -0.011  \pm  0.018$ \\ 
 99727  & 56644.842 & 650 &$  -0.072  \pm  0.019$ \\ 
\hline                                           
 101082 & 56643.813 & 200 &$ -14.069  \pm  0.014$ \\ 
 101166 & 56643.820 & 200 &$ -13.747  \pm  0.014$ \\ 
\hline                                           
 101916 & 56643.827 & 120 &$ -53.457  \pm  0.015$ \\ 
 101932 & 56643.832 & 180 &$ -53.118  \pm  0.016$ \\ 
\hline                                           
 110419 & 56645.804 & 600 &$ -15.719  \pm  0.204$ \\ 
 110433 & 56645.816 & 700 &$ -14.754  \pm  0.245$ \\ 
\hline                                           
 112222 & 56643.841 & 120 &$  -5.597  \pm  0.033$ \\ 
 112354 & 56643.846 & 350 &$  -1.038  \pm  0.047$ \\ 
\hline                                           
 112970 & 56643.863 & 200 &$   2.348  \pm  0.146$ \\ 
 112946 & 56643.870 & 250 &$   1.758  \pm  0.025$ \\ 
\hline                                           
 113579 & 56644.854 & 450 &$   7.022  \pm  0.041$ \\ 
 113597 & 56644.863 & 800 &$  52.181  \pm  0.113$ \\ 
\hline                                           
 117573 & 56643.893 & 300 &$  14.113  \pm  0.110$ \\ 
 117573 & 56645.862 & 600 &$  13.951  \pm  0.125$ \\ 
 117733 & 56643.900 & 350 &$  15.314  \pm  0.020$ \\ 
 117733 & 56645.895 & 700 &$  15.346  \pm  0.023$ \\ 
\hline                                           
 118254 & 56643.909 & 300 &$  30.368  \pm  0.016$ \\ 
 118251 & 56643.916 & 350 &$  29.621  \pm  0.014$ \\ 
\end{longtable} 
\end{center}

For 14 pairs (28 stars) two spectra were obtained at different epochs to
confirm the stability of our measurements and/or possibly identify
stars that, for whatever reason, do have variable radial velocity.  It
was found that about half of these stars indeed have variable velocity well
above statistical uncertainties. The other half shows remarkable
stability, with mean velocity difference between repeated measurements
of $62 \pm 9$ m/s, confirming the good quality of the data.\\ 
\indent
Finally, radial velocity standard stars were observed to check the
radial velocity zero point of the instrument. Specifically, we
obtained spectra of HD38230 and HD50692 twice in two consecutive
nights, finding consistent radial velocity within few tens of m/s.
There is, however, a statistically significant offset of $378 \pm 46$
m/s in the zero point when compared with catalogue values.  While
we cannot find a simple explanation for this offset, our main result
remain unaffected by this, because we are only interested in
$differences$ of radial velocity.

\section{Results}

\begin{figure*}[htbp]
\centering
\includegraphics[width=0.49\columnwidth]{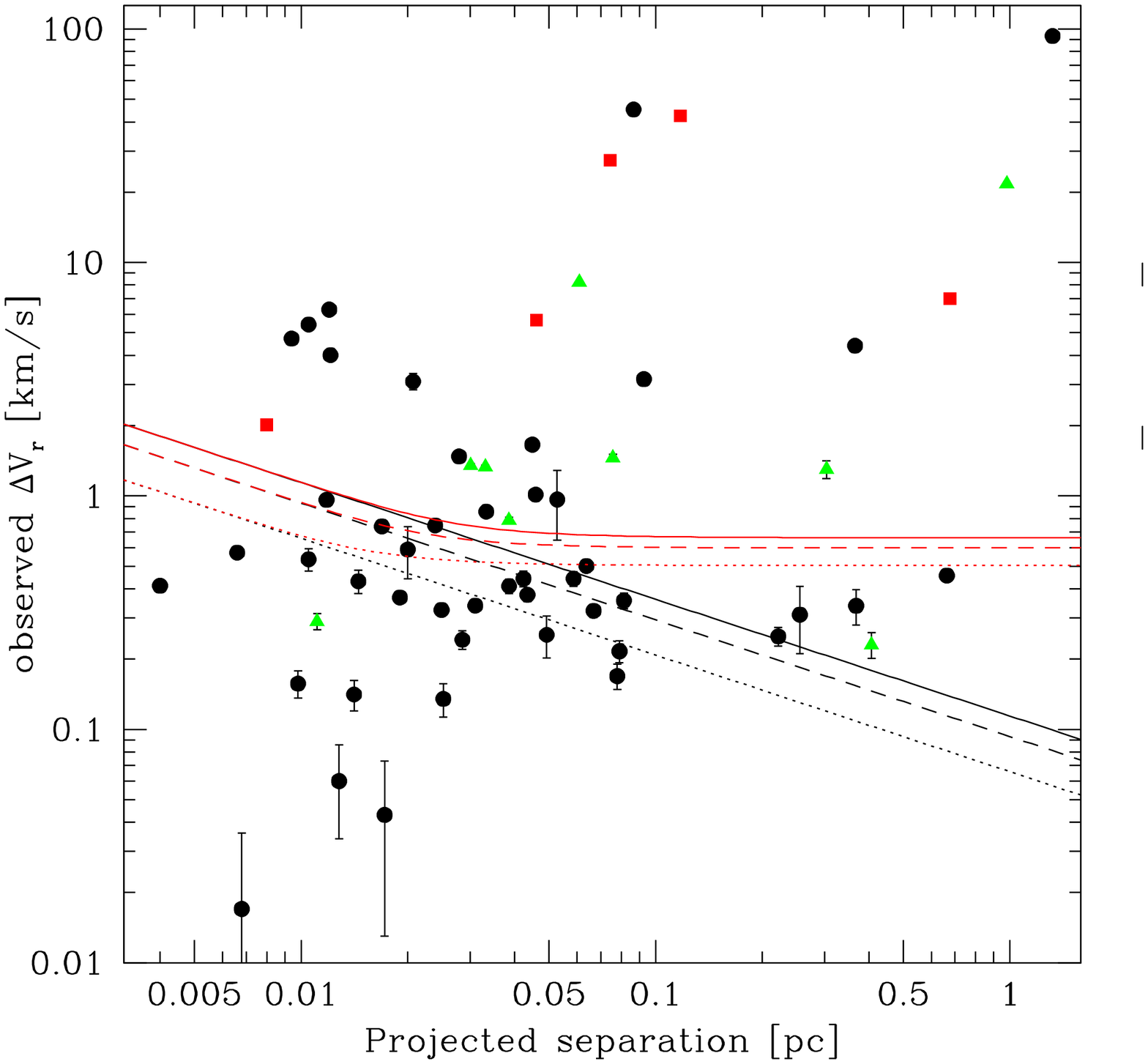}
\includegraphics[width=0.49\columnwidth]{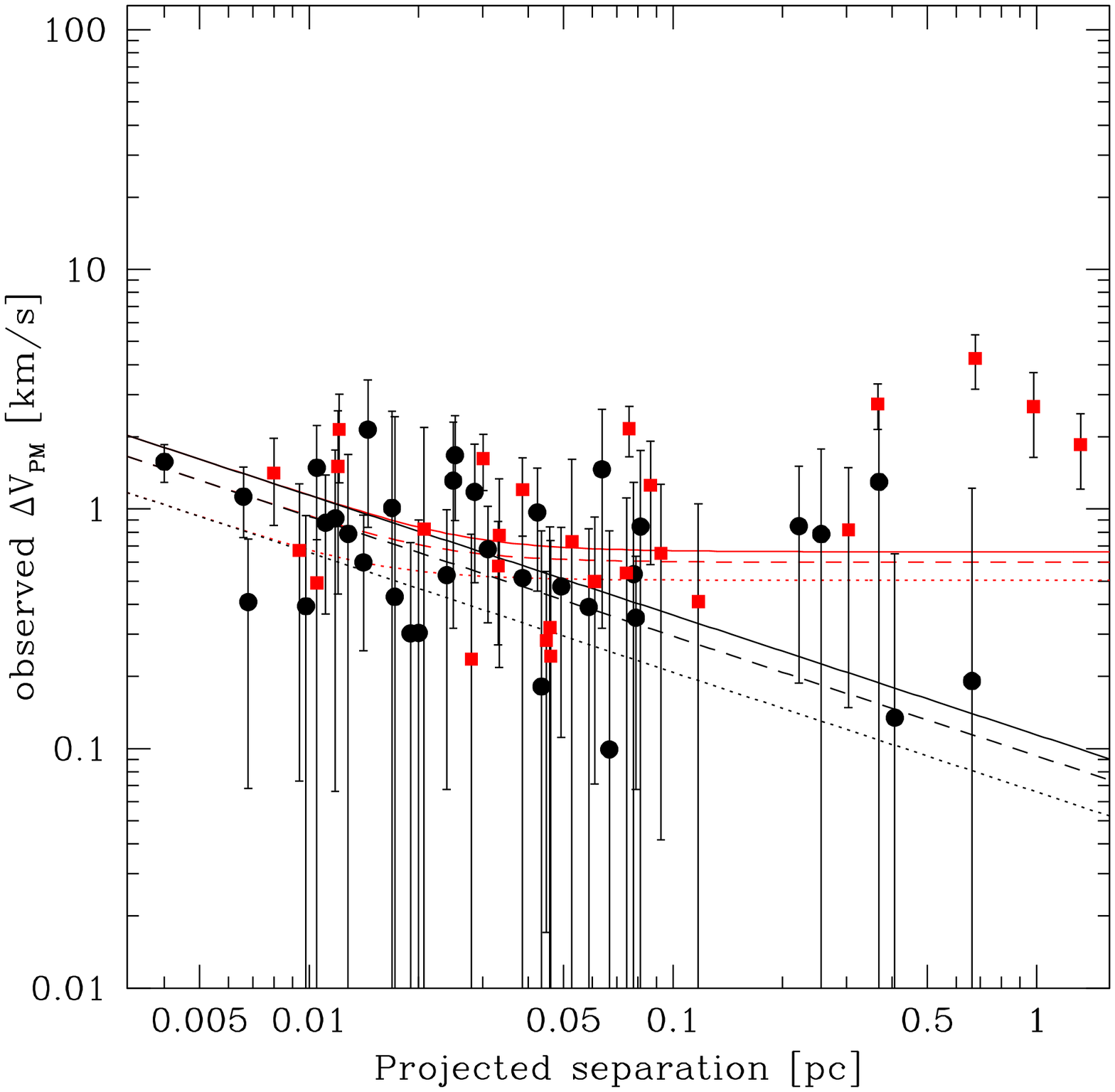}
\caption{ The measured radial velocity difference as a function of the
  projected separation for the observed pairs.  The black lines
  represent the relationship between total $\Delta V$ (not just the
  radial or tangential component) and star separation for binary stars
  with total mass of 1.0 (dotted line), 2.0 (dashed line) and 3.0
  (filled line) solar masses.  
  {\bf Left:} The radial component of the
  observed velocity $\Delta V_{r}$ difference is twice the component
  of the orbital velocity. Error bars give the 1$\sigma$ uncertainty,
  when not visible the error bar is smaller than the symbol.  Stars
  observed twice are marked as red squares if $\Delta V_{r}$ did vary
  more than 3$\sigma$ between observations, as triangles if $\Delta
  V_{r}$ remained constant within statistical errors.  
  {\bf Right:}
  The velocity difference along the tangential direction as derived
  from proper motions (filled circles).  Note the error bars are
  significantly larger than those for radial velocity
  measurements. Points are black or red according to whether $\Delta
  V_{r}$ is below or above the MOND prediction for a three solar
  masses star pair, respectively.  }\label{fig:rvpd}
\end{figure*}

The observed radial velocity difference $\Delta V_{r}$ for our 60
pairs is reported in Table \ref{Tab:3} and shown in Fig.\ref{fig:rvpd}
as a function of projected separation. For comparison, the tangential
component $\Delta V_{PM}$ from Hipparcos proper motions is also shown.
The difference between the two plots is striking, with approximately
50\% of the pairs having radial velocity clearly
incompatible with the pairs being bound systems (in any reasonable
scenario).  It is clear, however, that a bias is present because stars
with large proper motion difference would have not entered in the
Shaya \& Olling catalog in the first place. Therefore, adding accurate
radial velocities give us immediately the possibility to show that
either something alter the velocity of the stars, or many systems are
unbound in spite of the high probability to be bound assigned solely
according to proper motion data.

\begin{center}
\begin{longtable}{cccccc}
\caption{Table of difference in radial velocities.\\
Legend:\\
\textbf{NH1} = Number of Hipparcos catalogue of the first star;\\
\textbf{NH2} = Number of Hipparcos catalogue of the second star;\\
\textbf{sep} = Projected separation. The value is obtained from the angular separation multiplied by the weighted average of the parallaxes distances of the binary sistem;\\
\textbf{dpm} = Difference  in proper motion. The value is obtain combining the value of proper motion along two different direction;\\
\textbf{dRV} = Difference in radial velocity.\\
\textbf{$\Delta t$} = time interval between repeated observations in days}
\label{Tab:3}\\

\hline
NH1 & NH2 & sep & dpm & dRV & $\Delta t$\\
 &  & [pc] & [msec/yr] & [km/s] & [Days]\\                     
\hline
\endfirsthead

\multicolumn{5}{l}{{\bfseries \tablename\ \thetable{} -- continued from previous page}} \\
\hline
NH1 & NH2 & sep & dpm & dRV & $\Delta t$\\
 &  & [pc] & [msec/yr] & [km/s] & [Days]\\                     
\hline
\endhead

\hline
\multicolumn{6}{r}{ \bfseries Continued on next page}
\endfoot

\hline
\endlastfoot

$ 185 $&$ 190 $&$ 0.0249   \pm   0.0013 $&$ 3.360   \pm   1.800  $&$ 0.325   \pm   0.021 $\\ 
\hline                                    
$ 201 $&$ 206 $&$ 0.2554   \pm   0.0174 $&$ 1.726   \pm   1.597  $&$ 0.310   \pm   0.099 $\\ 
\hline                                    
$ 1891 $&$ 1887 $&$ 0.0121   \pm   0.0007 $&$ 5.805   \pm   1.656  $&$ -4.007   \pm   0.024 $\\ 
\hline                                    
$ 2292 $&$ 2350 $&$ 0.222   \pm   0.0084 $&$ 3.360   \pm   1.865  $&$ 0.251   \pm   0.023 $\\ 
\hline                                    
$ 4702 $&$ 4833 $&$ 0.6633   \pm   0.0335 $&$ 0.500   \pm   1.900  $&$ -0.456   \pm   0.018 $\\ 
\hline                                    
$ 8497 $&$ 8486 $&$ 0.0207   \pm   0.0001 $&$ 7.596   \pm   11.135  $&$ 3.091   \pm   0.250 $\\ 
\hline                                    
$ 10321 $&$ 10339 $&$ 0.0791   \pm   0.0008 $&$ 2.802   \pm   1.600  $&$ 0.215   \pm   0.023 $\\ 
\hline                                    
$ 11137 $&$ 11134 $&$ 0.0098   \pm   0.0006 $&$ 1.432   \pm   1.400  $&$ -0.157   \pm   0.021 $\\ 
\hline                                    
$ 11783 $&$ 11759 $&$ 0.0449   \pm   0.0003 $&$ 2.184   \pm   1.493  $&$ 1.658   \pm   0.028 $\\ 
\hline                                    
$ 15304 $&$ 15310 $&$ 0.0333   \pm   0.0011 $&$ 3.734   \pm   1.900  $&$ -0.856   \pm   0.027 $\\ 
\hline                                    
$ 15527 $&$ 15526 $&$ 0.0435   \pm   0.0009 $&$ 1.082   \pm   2.596  $&$ -0.377   \pm   0.024 $\\ 
\hline                                    
$ 17118 $&$ 17126 $&$ 0.0094   \pm   0.0002 $&$ 3.413   \pm   2.128  $&$ 4.718   \pm   0.182 $\\ 
\hline                                    
$ 19335 $&$ 19255 $&$ 0.0758   \pm   0.0003 $&$ 22.064   \pm   3.733  $&$ -1.385   \pm   0.056 $&$ 1.1$\\ 
$ 19335 $&$ 19255 $&$ - $&$ -  $&$  -1.532   \pm   0.058 $\\ 
\hline                                    
$ 19859 $&$ 19855 $&$ 0.0066   \pm   0.0001 $&$ 11.226   \pm   2.139  $&$ 0.572   \pm   0.020 $\\ 
\hline                                    
$ 20342 $&$ 20338 $&$ 0.0105   \pm   0.0005 $&$ 9.055   \pm   3.200  $&$ -10.462   \pm   0.059 $\\ 
\hline                                    
$ 20669 $&$ 20637 $&$ 0.0638   \pm   0.0027 $&$ 4.880   \pm   2.700  $&$ 0.501   \pm   0.031 $\\ 
\hline                                    
$ 21537 $&$ 21534 $&$ 0.0252   \pm   0.0009 $&$ 5.324   \pm   1.701  $&$ -0.135   \pm   0.022 $\\ 
\hline                                    
$ 21704 $&$ 21702 $&$ 0.0172   \pm   0.0011 $&$ 1.020   \pm   3.396  $&$ -0.043   \pm   0.030 $\\ 
\hline                                    
$ 22611 $&$ 22604 $&$ 0.0285   \pm   0.0006 $&$ 4.254   \pm   1.725  $&$ -0.241   \pm   0.022 $\\ 
\hline                                    
$ 24046 $&$ 24035 $&$ 0.0609   \pm   0.0016 $&$ 2.640   \pm   1.600  $&$ 8.223   \pm   0.032 $&$ 1.0$\\ 
$ 24046 $&$ 24035 $&$ - $&$ -  $&$  8.226   \pm   0.033 $\\ 
\hline                                    
$ 25278 $&$ 25220 $&$ 0.0493   \pm   0.0003 $&$ 7.033   \pm   3.468  $&$ -0.253   \pm   0.052 $\\ 
\hline                                    
$ 33705 $&$ 33691 $&$ 0.0587   \pm   0.0007 $&$ 2.209   \pm   1.799  $&$ -0.441   \pm   0.032 $\\ 
\hline                                    
$ 34426 $&$ 34407 $&$ 0.0386   \pm   0.001 $&$ 5.482   \pm   1.400  $&$ 0.757   \pm   0.026 $&$ 319.1$\\ 
$ 34426 $&$ 34407 $&$ - $&$ -  $&$  0.812   \pm   0.022 $\\ 
\hline                                    
$ 34714 $&$ 34700 $&$ 0.4069   \pm   0.0091 $&$ 0.608   \pm   1.603  $&$ -0.294   \pm   0.029 $&$ 321.1$\\ 
$ 34714 $&$ 34700 $&$ - $&$ -  $&$  -0.168   \pm   0.025 $\\ 
\hline                                    
$ 39457 $&$ 39452 $&$ 0.0169   \pm   0.0012 $&$ 2.955   \pm   3.200  $&$ -0.741   \pm   0.022 $\\ 
\hline                                    
$ 40918 $&$ 40882 $&$ 0.0461   \pm   0.0012 $&$ 1.393   \pm   2.000  $&$ 5.661   \pm   0.025 $&$ 321.1$\\ 
$ 40918 $&$ 40882 $&$ - $&$ -  $&$  3.204   \pm   0.021 $\\ 
\hline                                    
$ 44520 $&$ 44518 $&$ 0.008   \pm   0.0005 $&$ 5.738   \pm   1.600  $&$ -2.013   \pm   0.022 $&$ 321.0$\\ 
$ 44520 $&$ 44518 $&$ - $&$ -  $&$  0.165   \pm   0.024 $\\ 
\hline                                    
$ 44858 $&$ 44864 $&$ 0.0111   \pm   0.0004 $&$ 4.123   \pm   1.723  $&$ -0.268   \pm   0.023 $&$ 321.2$\\ 
$ 44858 $&$ 44864 $&$ - $&$ -  $&$  -0.312   \pm   0.021 $\\ 
\hline                                    
$ 45811 $&$ 45802 $&$ 0.0745   \pm   0.0014 $&$ 1.746   \pm   1.300  $&$ -11.640   \pm   0.092 $&$ 319.1$\\ 
$ 45811 $&$ 45802 $&$ - $&$ -  $&$  27.428   \pm   0.072 $\\ 
\hline                                    
$ 45836 $&$ 45859 $&$ 0.031   \pm   0.0004 $&$ 4.386   \pm   1.684  $&$ -1.353   \pm   0.021 $&$ 321.2$\\ 
$ 45836 $&$ 45859 $&$ - $&$ -  $&$  -1.316   \pm   0.021 $\\ 
\hline                                    
$ 47436 $&$ 47403 $&$ 0.1174   \pm   0.0019 $&$ 1.769   \pm   1.947  $&$ 40.506   \pm   0.035  $&$ 321.0$\\ 
$ 47436 $&$ 47403 $&$ - $&$ -  $&$  42.465   \pm   0.035 $\\ 
\hline                                    
$ 50325 $&$ 50327 $&$ 0.0279   \pm   0.0007 $&$ 0.943   \pm   1.529  $&$ -1.477   \pm   0.035 $\\ 
\hline                                    
$ 52787 $&$ 52776 $&$ 0.0386   \pm   0.0009 $&$ 3.245   \pm   3.031  $&$ -0.410   \pm   0.029 $\\ 
\hline                                    
$ 54692 $&$ 54681 $&$ 0.0309   \pm   0.0007 $&$ 7.203   \pm   1.300  $&$ 1.381   \pm   0.021 $&$ 321.2$\\ 
$ 54692 $&$ 54681 $&$ - $&$ -  $&$  1.324   \pm   0.020 $\\ 
\hline                                    
$ 57858 $&$ 57856 $&$ 0.0145   \pm   0.0007 $&$ 4.640   \pm   2.000  $&$ -0.430   \pm   0.050 $\\ 
\hline                                    
$ 58067 $&$ 58073 $&$ 0.0141   \pm   0.0005 $&$ 3.206   \pm   1.300  $&$ 0.141   \pm   0.021 $\\ 
\hline                                    
$ 58751 $&$ 58722 $&$ 0.0927   \pm   0.0028 $&$ 2.417   \pm   1.600  $&$ 3.163   \pm   0.046 $\\ 
\hline                                    
$ 64057 $&$ 64059 $&$ 0.0068   \pm   0.0002 $&$ 2.377   \pm   1.400  $&$ 0.017   \pm   0.019 $\\ 
\hline                                    
$ 64444 $&$ 64443 $&$ 0.0105   \pm   0.0008 $&$ 1.526   \pm   2.100  $&$ -5.419   \pm   0.034 $\\ 
\hline                                    
$ 65602 $&$ 65574 $&$ 0.0459   \pm   0.0009 $&$ 2.267   \pm   2.632  $&$ -1.014   \pm   0.025 $\\ 
\hline                                    
$ 65899 $&$ 65884 $&$ 1.3181   \pm   0.0714 $&$ 5.292   \pm   1.300  $&$ -93.313   \pm   0.024 $\\ 
\hline                                    
$ 66749 $&$ 66717 $&$ 0.6772   \pm   0.0297 $&$ 10.508   \pm   1.900  $&$ -6.546   \pm   0.197 $&$ 1.0$\\ 
$ 66749 $&$ 66717 $&$ - $&$ -  $&$  -6.996   \pm   0.066 $\\ 
\hline                                    
$ 67250 $&$ 67041 $&$ 0.9788   \pm   0.0298 $&$ 5.869   \pm   1.600  $&$ 21.647   \pm   0.018 $&$ 1.0$\\ 
$ 67250 $&$ 67041 $&$ - $&$ -  $&$  21.617   \pm   0.018 $\\ 
\hline                                    
$ 68830 $&$ 68833 $&$ 0.012   \pm   0.0008 $&$ 4.002   \pm   2.000  $&$ -6.279   \pm   0.280 $\\ 
\hline                                    
$ 69701 $&$ 69962 $&$ 0.3682   \pm   0.0016 $&$ 12.462   \pm   9.283  $&$ 0.338   \pm   0.058 $\\ 
\hline                                    
$ 71726 $&$ 71737 $&$ 0.078   \pm   0.0021 $&$ 2.159   \pm   2.195  $&$ -0.170   \pm   0.021 $\\ 
\hline                                    
$ 74442 $&$ 74439 $&$ 0.0118   \pm   0.0007 $&$ 2.596   \pm   1.615  $&$ -0.962   \pm   0.023 $\\ 
\hline                                    
$ 74666 $&$ 74674 $&$ 0.019   \pm   0.0001 $&$ 1.720   \pm   1.700  $&$ -0.367   \pm   0.021 $\\ 
\hline                                    
$ 85620 $&$ 85575 $&$ 0.0424   \pm   0.0007 $&$ 4.528   \pm   1.700  $&$ -0.442   \pm   0.034 $\\ 
\hline                                    
$ 96895 $&$ 96901 $&$ 0.004   \pm   0 $&$ 15.728   \pm   1.614  $&$ -0.412   \pm   0.016 $\\ 
\hline                                    
$ 97295 $&$ 97222 $&$ 0.0814   \pm   0.0004 $&$ 8.443   \pm   5.466  $&$ -0.355   \pm   0.027 $\\ 
\hline                                    
$ 99729 $&$ 99727 $&$ 0.0128   \pm   0.0007 $&$ 2.717   \pm   2.200  $&$ 0.061   \pm   0.026 $\\ 
\hline                                    
$ 101082 $&$ 101166 $&$ 0.0669   \pm   0.0009 $&$ 0.316   \pm   1.600  $&$ -0.322   \pm   0.020 $\\ 
\hline                                    
$ 101916 $&$ 101932 $&$ 0.031   \pm   0.0003 $&$ 4.880   \pm   1.749  $&$ -0.339   \pm   0.022 $\\ 
\hline                                    
$ 110419 $&$ 110433 $&$ 0.0528   \pm   0.0025 $&$ 1.924   \pm   1.603  $&$ -0.965   \pm   0.319 $\\ 
\hline                                    
$ 112222 $&$ 112354 $&$ 0.3654   \pm   0.0178 $&$ 13.788   \pm   2.100  $&$ -4.559   \pm   0.057 $\\ 
\hline                                    
$ 112970 $&$ 112946 $&$ 0.02   \pm   0.0003 $&$ 1.082   \pm   1.500  $&$ 0.590   \pm   0.148 $\\ 
\hline                                    
$ 113579 $&$ 113597 $&$ 0.0867   \pm   0.0019 $&$ 8.645   \pm   3.234  $&$ -45.159   \pm   0.120 $\\ 
\hline                                    
$ 117573 $&$ 117733 $&$ 0.3035   \pm   0.0079 $&$ 3.023   \pm   1.769  $&$ -1.201   \pm   0.112 $&$ 2.0$\\ 
$ 117573 $&$ 117733 $&$ - $&$ -  $&$  -1.395   \pm   0.127 $\\ 
\hline                                    
$ 118254 $&$ 118251 $&$ 0.0239   \pm   0.0005 $&$ 2.594   \pm   1.600  $&$ 0.747   \pm   0.021 $\\ 
\hline                                    
                               
\end{longtable} 
\end{center}

The Newtonian and modified gravity upper limit velocity expectations
are also reported.  Note in our sample $3M_{\odot}$ is the very
maximum $total$ mass a pair can have, thus, if all pairs were bound
all points should fall, as a minimum, below the modified dynamics
prediction.  As mentioned before, for 14 pairs (28 stars) we have
repeated measurements. Scanning the velocity for these stars
(Tab. \ref{Tab:2}), we identify 3 possible cases: the velocity of both
stars in the pair varied, only one varied, neither varied. Here it is
also important to take into account the time separation between
observations, in our case either about 1-2 days or $\sim 300$
days.\\ \indent For instance, pair 66749/66717, which was observed at
1 day distance, showed a substantial variation of the radial velocity
of both stars (600 and 100 m/s).  The most probable explanation for
this variation in such a short amount of time is that both stars are
them self double (we are then dealing with a quadruple system) of
which we are mapping the orbital velocity of two close internal binary
systems.  While pairs like this are still potentially useful for
testing Newtonian dynamics, they require a substantial observational
effort in order to average out these internal orbital variations. And
the situation would be even worse if the internal systems are
moderately wide, because in this case observations should cover a much
longer period of time before velocity variation can be identified.  We
feel most of the points above the maximum theoretical limit belong to
this class.\\

\indent In pairs like 47436/47403 and 45811/45802 only one of the two stars
have variable radial velocity. Again, the simplest explanation is that
we are dealing with a triple system. Star 45802 is particularly
striking, showing a change of about 40 km/s over a baseline of 319
days. Possibly, if observed at 1 day distance, this star will have
shown a variation similar to the one of 66749. Also in this case a
substantial observational effort is required to wash out the
variations.\\

\indent Finally, pairs 34426/34407, 44858/44864, 45836/45859, 54692/54681
observed over a large time baseline, do show constant velocity (within
statistical uncertainties).  These are the most interesting
cases. Unfortunately we have only four of them (the remaining 5 pairs
with stable velocity were observed at short time distance so the
result is less significant).

Of the 50 pairs with projected separation $< 0.15$ pc, where Newtonian
and modified dynamics made very similar predictions, 28 pairs
are below the Newtonian upper limit for $3M_{\odot}$ pair (and one more
below the modified gravity limit).

In view of what we found for the stars with repeated observations, a
number of the systems below the Newtonian limit could be
multiple. While it is possible that by chance these pairs have been
observed in an orbital phase corresponding to a low velocity
difference, we believe that this is unlikely.
So we are confident that a substantial fraction of these 28 pairs
are bound systems suitable for testing Newtonian dynamics. 

Considering the remaining 10 pairs with separation $>0.15$pc one
immediately notices the absence of systems below the Newtonian limits,
while there are 5 pairs formally in agreement with the modified
dynamics limit. The remaining 5 pairs have radial velocity $>1$ km/s
and must be either unbound or multiple systems.  The 5 pairs sitting
in the region between the two models are unquestionably the most
interesting and certainly deserve further investigation.


\begin{figure}[h]
 \centering
 \includegraphics[width=0.49\columnwidth]{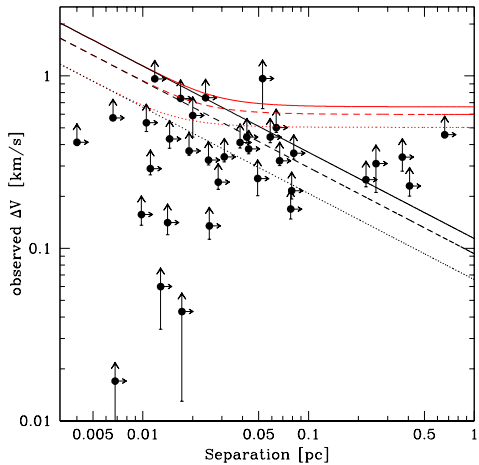}

 \caption{The observed velocity difference vs the pair    
    separation S for pairs with $\Delta V_{r} <$ 2 km/s. 
   The observed separation corresponds to a lower limit to the true one.
   A few points beyond 0.15    pc separations might correspond to 
   gravitationally bound systems.
 }\label{fig:fullDV}
\end{figure}

\section{Conclusions}
\noindent

We have shown that for about half of the observed pairs, accurate radial velocity measurements
 supports the
hypothesis that these pairs are indeed gravitationally bound systems.
On the other hand the radial velocity data also show that about 50\%
of the pairs cannot be bound systems, making clear that proper motion
data alone is not sufficient to perform the proposed test
(Ref.~\refcite{Hernandez12}).  Moreover, repeated radial velocity
measurements appear to be mandatory in order to further eliminate
multiple systems of stars that might have escaped detection in the proper motion
data analysis.

With the present dataset, we can only point out that looking at the
wider pairs, at least a few of them seem at odd with Newtonian
prediction while remaining formally consistent with modified dynamics
(see Figure \ref{fig:fullDV}).  Since we are able to observe
accurately only one component of the velocity difference and the real
separation between the stars will be larger than that observed (namely
the projected one), some additional pairs would probably move into
this intermediate region when these effects will be taken into account
(see Figure \ref{fig:fullDV}).  While it is premature from this
data to draw any conclusion about dynamics at low
accelerations, it is worth to underline that our observations suggest
the existence of very wide gravitationally bound binary stars in the
solar neighborhood, making the proposed test feasible.

In the near future, the GAIA satellite (http://sci.esa.int/gaia/) will
provide proper motion data with very high accuracy, also dramatically
increasing the number of stars for which such data will be
available. GAIA will provide limited radial velocity information. Still,
these radial velocity data, while of
insufficient quality for the proposed test, will prove extremely
valuable for selecting wide pair candidates, dramatically reducing the
number of false positive.  Once selected according to proper motion
and radial velocity, a minimum of two high resolution spectra with
radial velocity accuracy $\sim 100$ m/s will have to be obtained for
each component of the pair, in this way pinning down the third
component of the velocity vector and eliminating stars that for
whatever reasons have variable radial velocity.  Our results suggest
that once all these new data will be available, a sufficiently large
number of extremely wide binaries will be identified, providing enough
information to build the ``rotation curve'' of binary stars up to
$\sim 1$ pc separation.  This will probe in the cleanest possible way
Newtonian dynamics, providing an extremely powerful tool to address one
of the most puzzling problem that modern astrophysics is facing: the
true nature of the dark matter.

\end{document}